\documentclass[12pt]{article}
\usepackage{amsmath}
\usepackage{amssymb}
\usepackage{amsthm}
\usepackage{rotating}
\usepackage{bm}
\usepackage{multirow}
\usepackage[utf8]{inputenc}

\usepackage{graphicx}
\usepackage{enumerate}
\usepackage{natbib}
\usepackage{url} 
\usepackage{color, colortbl}
\usepackage{xcolor}
\usepackage{soul}
\usepackage{float}

\definecolor{Gray}{gray}{0.9}

\newcommand{\black}{\color{black}}

\newcommand{\be}{\begin{eqnarray*}}
	\newcommand{\ee}{\end{eqnarray*}}
\newcommand{\ben}{\begin{eqnarray}}
\newcommand{\een}{\end{eqnarray}}

\newcommand{\g}{\,\vert\,}
\newcommand{\hthetao}{\hat{\theta}_o}
\newcommand{\hthetar}{\hat{\theta}_r}

\newtheorem{result}{Result}
\newtheorem*{result*}{Result}
\newtheorem{proposition}{Proposition}
\newtheorem*{lemma*}{Lemma}

\begin{document}

\def\spacingset#1{\renewcommand{\baselinestretch}%
{#1}\small\normalsize} \spacingset{1}


  \title{\bf Assessing replication success via skeptical mixture priors\\
  }
  \author{Guido Consonni\thanks{
    Department of Statistical Sciences, Universit\`{a} Cattolica del Sacro Cuore, Milano}
    \ and Leonardo Egidi\thanks{Department of Economics, Business, Mathematics, and Statistics \emph{Bruno de Finetti}, University of Trieste}}
    
  \maketitle


\bigskip
\begin{abstract}

There is a growing interest in the analysis of  replication studies
of  original findings
across many disciplines.
 When  testing a hypothesis for an effect size,
two  Bayesian approaches stand out for their principled use of
the Bayes factor (BF),  namely
the {replication} BF and the {skeptical} BF.
In particular, the latter BF
is based on the skeptical prior, which represents the
 opinion of an investigator who is unconvinced  by the
original findings and wants to challenge them.
We embrace the skeptical perspective, and   elaborate  a novel mixture prior which
incorporates   skepticism while at the same time  controlling for  prior-data conflict within the original data.
Consistency properties  of the resulting skeptical mixture BF are provided
together with an extensive   analysis of the main features of our proposal.
Finally, we  apply our methodology   to
data from the Social Sciences Replication Project.
In particular we show that,
for some  case studies
where prior-data conflict is an issue,
our method uses a more realistic  prior and leads to
evidence-classification  for replication success
which differs from the  standard skeptical approach.
\end{abstract}
\black

\noindent%

{\it Keywords:} 
Bayes factor, Bayesian hypothesis testing, consistency, prior-data conflict, replication studies, reverse-Bayes
\vfill

\newpage
\spacingset{1.9} 
\section{Introduction and Background}
\label{sec:intro}

\color{black}

The so-called \lq \lq replication crisis\rq \rq{}  has been plaguing scientific research across many disciplines \citep{open2015estimating,  camerer2018evaluating}, and this
has generated a growing interest in  the analysis of replication studies.
Several attempts have been carried out to pinpoint the notion of  \lq \lq replication success\rq \rq{};  see for instance
 \cite{hutton2020discussion}, \citet{anderson2016there},
\cite{johnson:etal:2017:On:the:Reproducibility},
\cite{Ly:etal:2018:replication:bayes:factor},
\cite{hedges2019statistical},
\cite{harms2019bayes} and
\citet{held2020new}.

Within the Bayesian framework,  the Bayes Factor (BF)
has played an important role in evaluating the evidence  for replication success. Notable examples are the
\emph{replication} BF  \citep{verhagen2014bayesian} and  the  skeptical BF \citep{pawel2022sceptical}.
In this paper we analyze consistency issues  for both  BF's, and then discuss potential prior-data conflict underlying the use of the skeptical BF. This investigation will eventually lead to our proposal which in a loose sense
can be regarded as a natural halfway house between the two.
\black

We review the main features of the replication BF in Section 1.1 and of the skeptical BF in Section 1.2.
Sections 1.3 to 1.5 contain the motivation for our work and present useful technical details.
In particular,
Section 1.3 is devoted to consistency issues under replication data; specifically, consistency of the replication BF is proven, while this is not the case for the skeptical BF.
A more specific notion of consistency, named information consistency, is dealt with in Section 1.4: again the replication BF satisfies it; on the other hand we argue that information
consistency cannot be properly formulated when it comes to the skeptical BF because the two contrasting hypotheses are not nested.
Finally,  Section 1.5  highlights  why   {prior-data conflict} should be taken into consideration in replication studies, and   provides some background on the topic.
 Section 2 presents our proposal, namely   the skeptical mixture prior with the allied  BF, along with a result on consistency and the extensive analysis of an example  (Section 2.1).
Section 3 contains an  application to  case studies
from the \emph{Social Sciences Replication Project} reported in \citet{camerer2018evaluating},
together
with an assessment  of our method, also from a comparative viewpoint.
Some issues worthy of discussion are then summarized in Section 4.
To ease the flow of ideas, several technical details have been collected in the  Supplementary material.

\subsection{The Replication Bayes Factor}

Consider two Bayesian models for the same observable $y$
\ben
\label{eq: Hj-def}
H_j: \{ f(y \g H_j, \theta_j); \, f(\theta_j \g H_j) \}, \,  j=1,2,
\een
where $f(y \g H_j, \theta_j)$ is the data  distribution under $H_j$ indexed by parameter $\theta_j$, and   $f( \theta_j \g H_j)$ is the corresponding parameter prior.
We evaluate the plausibility of $H_1$ relative to $H_2$ based on data $y$ through the Bayes Factor (BF)
\ben
BF_{1:2}(y)=\frac{f(y\g H_1) }{f(y\g H_2)},
\een
where
$f(y\g H_j)=\int f(y \g H_j, \theta_j)f(\theta_j \g H_j)d\theta_j$ is the marginal data distribution under $H_j$, also named the marginal likelihood of $H_j$.

In Equation \eqref{eq: Hj-def}  both the data distribution and the prior
may depend on $H_j$. In our setting however,  the family of data distributions is the same under $H_1$ and $H_2$ with the same parameter $\theta$ say, so that
$f(y \g H_j, \theta_j)=f(y \g  \theta)$; as a consequence  $H_j$ characterizes only  the prior for $\theta$, and model comparison reduces to a Bayesian hypothesis testing problem.

Let $\theta$ be   the effect of a treatment on an outcome of  interest,
and
 $\hthetao$ and  $\hthetar$ denote estimators (typically MLE)  of $\theta$ obtained under the \emph{original} and the \emph{replication} study respectively, with corresponding standard errors $\sigma_o$ and $\sigma_r$. Following common practice in meta-analytic studies,  we further assume that the sample sizes $n_k$ are sufficiently large to justify a normal  distribution for the estimators, so that  $\hat{\theta}_k \g \theta \sim N(\theta, \sigma_k^2)$ with $\sigma_k$ known, $k \in \{ o,r \}$. This represents a reasonable approximation for various types of effect sizes,   including  means and mean differences,    odds ratios, hazard ratios, risk ratios  or correlation coefficients, usually after a suitable transformation; see for instance
 \citet[Section 2.4]{Spiegelhalter:et:al:book:2003}.

The following notation  will also be useful in the sequel. Denote the $z$-values associated to the estimates of the two studies with $z_o=\hthetao/\sigma_o$,  $z_r=\hthetar/\sigma_r$, respectively; the relative effect
estimate with $d=\hthetar/\hthetao$; the variance ratio with $c=\sigma_o^2/\sigma_r^2$.
Since for many  types of effect sizes the variances are inversely
proportional to the sample size, often one can safely assume that $\sigma_k^2=\sigma^2/n_k$, $k \in \{ o,r \}$,
where  $\sigma^2$ is the unitary variance in each study. In this case
$c = n_r/n_o$, the ratio between the replication and the original sample size.

Of particular interest in our setting is the situation wherein  $z_o$ is sufficiently large in absolute value, so  that the original experiment is believed to provide substantial evidence that there is truly an effect, i.e. $\theta \neq 0$.
  To evaluate to what extent a replication study resulted in  a success, thus  confirming the original finding,  \citet{verhagen2014bayesian} compared two hypotheses
using  replication data $\hat{\theta}_r \g \theta \sim N(\theta,
\sigma_r^2)$.
The first one is the standard null hypothesis
$
H_0: \theta=0
$
of no effect. The second one  reflects the opinion of an \emph{advocate} who believes
the effect to be consistent with that found in the original study. This is quantified through a posterior distribution on $\theta$, conditionally on the original data $\hthetao$, and based on
 a flat prior for $\theta$.   The resulting \emph{advocate prior} becomes
\be
H_A: \theta \sim N(\hthetao, \sigma_o^2).
\ee
The BF of $H_0$ against $H_A$,
\be
BF_{0:A}(\hthetar)=\frac{f(\hthetar \g H_0)}{f(\hthetar \g H_A)} \equiv BF_R,
\ee
is named
the \emph{Replication Bayes factor}.
It can be  verified that 
\begin{equation}
\label{eq:BR}
BF_R = \sqrt{1+c}\ \exp \left\{ -\frac{z_o^2}{2} \left(d^2 c-\frac{(1-d)^2}{1/c+1}  \right) \right\}.
\end{equation}

Replication success is declared whenever $BF_R$ is  sufficiently low to  provide convincing  evidence against $H_0$, based on conventional evidence thresholds essentially  dating back to \cite{Jeffreys:1961}; see for instance
\citet[Table 1]{schonbrodt2018bayes}.
It may be observed that the replication BF is a  \textit{partial} BF
\citep{o2004kendall} for  checking $H_0$ against its complement $\theta \neq 0$
when the prior under the latter is flat. In this context,
$\hthetar$ is used as  comparison data and   $\hthetao$ as training data; see also \cite{ly2019replication}.
Notice that $BF_R$ provides an answer to the following  question: \lq \lq In the replication experiment, is the effect absent or is it similar to what was found in the original one?\rq \rq{}, where
 the latter supposition is represented through $H_A$. This should be contrasted with more traditional default Bayesian testing methods,  where the alternative is usually  a relatively uninformative prior centered  on the null value $\theta=0$; see for instance    \cite{wetzels2012default}.
We highlight that  $BF_R$  establishes a useful connection between the replication and the original experiment, through the  advocate prior. However,
replication success is declared  on the basis of the evidential strength against $H_0$ when compared to $H_A$ \emph{solely} under the  {replication} data. In other words there is no explicit  consideration of the evidence  against $H_0$ provided by the
original data.
This issue in taken up in the next section.

\subsection{The Skeptical Bayes Factor}
\label{subsection:the-skeptical}

\citet{pawel2022sceptical} propose a different  route to establish replication success. Their key idea  is to compare  two particular BF's: one  based on the {original} data,   and the other one on the {replication} data.
For the former they  compare the standard  null hypothesis of no effect $H_0$ against that of a \emph{skeptic} who is unconvinced by the result.
This view is operationalized through a \emph{skeptical} Normal prior,  centered on zero and with a variance $\sigma_S^2=g {}\sigma_o^2$, where $g$ is chosen   so that the resulting BF provides  unconvincing  evidence against the null hypothesis.
In other words, the skeptic wishes to \lq \lq challenge\rq \rq{} the original finding, and requires  the Bayes factor to attain a value so that s/he cannot take a definitive commitment against the null, and thus further investigation (namely the replication experiment) is called for.
More formally, let $0<\gamma<1$ be a value such that
the comparison of
\ben
\label{eq:H-S}
H_0: \theta=0 \quad \mbox{vs} \quad H_S: \theta \sim N(0, g_\gamma \sigma_o^2)
\een
leads to
\be
BF_{0:S}(\hthetao; g_\gamma)=\gamma,
\ee
with the understanding that values above $\gamma$ would \emph{not} be  considered  adequate  evidence against $H_0$.
For instance   $\gamma=1/10$ could be a suitable choice, because values of  $BF_{0:S}$ in the bracket (1/10,1/3) provide only moderate evidence against $H_0$, while those in the bracket (1/30, 1/10)
    imply strong evidence against $H_0$; see again  Table 1 of   \citet{schonbrodt2018bayes}  (notice however that the BF in their table is  the reciprocal of ours). A slightly different classification scheme is available in \citet{Kass:Raftery:1995}.

We note that the skeptical prior $N(0, g_\gamma \sigma_o^2)$  is constructed through  a \emph{reverse-Bayes}
methodology, a technique
dating back to \citet{Good1950book}.
The reason is that the prior is specified  \emph{after} the data are in, because the BF is required to attain a specific value on the original data $\hthetao$; see
\cite{held-etal-2022reverse} for an insightful discussion of reverse-Bayes ideas.

The value of $g_\gamma$, whose dependence on the original data is omitted  for simplicity,   can be explicitly computed as in  \citet[formula (3)]{pawel2022sceptical}.
It must be pointed out however that $g_\gamma$ will not  exist when
 $BF_{0:S}(\hthetao, g)$
  is always above $\gamma$ for any $g>0$: \black this
 happens for instance when $|z_o|\leq 1$,
 and  $\gamma \leq 1$; but it may  also happen for  $1<|z_o|\leq 2$ if $\gamma$ is  smaller than  1/3.
  These  situations however are hardly of interest,  because  in both cases either no effect is claimed,  or the evidence of an effect is  not particularly strong, and so replication of the experiment is often not even considered.  On the other hand, when $BF_{0:S}=\gamma$ is attainable,  there will  typically be two values of $g_\gamma$  leading to this result. The higher value, which is usually much greater than the smaller value,  is merely an instance of the Jeffreys-Lindley paradox
\citep{Shafer1982Lindleysparadox}, and is accordingly discarded because it represents vagueness rather than  skepticism.

Clearly the skeptical prior is data-dependent; however its use is  confined to obtain a BF whose value, on the original data,  is set based on external considerations. The skeptical distribution will then be used as a  regular   prior to construct a BF
based on the
replication  data $\hthetar$, and in that context is \emph{not} data-dependent.

Turning to replication data, one then  compares the  skeptical prior $H_S:  \theta \sim N(0, g_\gamma \sigma_o^2)$ against the advocate $H_A: \theta \sim N(\hthetao, \sigma_o^2)$, leading to
\ben
\label{eq:BFSA}
BF_{S:A}(\hthetar; g_\gamma)=\sqrt{\frac{1/c+1}{1/c+g_\gamma}}
\exp \left \{ -\frac{z_o^2}{2} \left(    \frac{d^2}{1/c+g_\gamma}-\frac{(d-1)^2}{1/c+1} \right)   \right \},
\een
where $z_o=\hthetao/\sigma_o$,
$d=\hthetar/\hthetao$,
and $c=\sigma_o^2/\sigma_r^2$.
Replication success at level $\gamma$ is declared if
\ben
\label{eq:BFSAlessBF0S}
BF_{S:A}(\hthetar; g_\gamma) \leq BF_{0:S}(\hthetao; g_\gamma)=\gamma.
\een
In the words of \citet{pawel2022sceptical}
\lq \lq It is natural
to consider a replication successful if the replication data favor the advocate over the skeptic
to a higher degree than the skeptic's initial objection to the original study\rq\rq{}.

Rather than fixing a value $\gamma$, and then checking whether Equation \eqref{eq:BFSAlessBF0S} holds, one might instead look for the smallest $\gamma$ satisfying \eqref{eq:BFSAlessBF0S}, namely
\begin{align}
\label{eq:BF-S}
BF_S \equiv  \text{inf} \ \{\gamma: BF_{S:A}(\hat{\theta}_r; g_{\gamma}) \leq \gamma  \}.
\end{align}
The value $BF_S$ is called the \emph{skeptical BF}, and  represents the smallest $\gamma$ level for which replication success can be established. It may happen that $BF_S$ does not exist, because there is no $\gamma$ for which replication success can be established, but this usually occurs when $|z_o|$, or $|d|=|z_r|/|z_o|$,  or both are too small. More details are provided in \citet{pawel2022sceptical}.

\subsection{Consistency}
\label{subsection:consistency}

\emph{Model selection consistency} \citep{liang2008mixtures},  or simply  consistency,
is the property of a statistical  procedure
 to recover the true model (or hypothesis)  as the sample size grows.
Below we analyze separately the behavior of the replication  and the skeptical Bayes factor.
In both cases consistency is evaluated relative to
a sequence of
replication datasets whose  sample size is assumed to grow indefinitely.
\begin{proposition}
\label{prop:consistency-BR}
Consider  a sequence of replication datasets with increasing sample size $n_r=1, 2, \ldots$.
Assume  there exists a
corresponding sequence of estimators
$\{  \hat{\theta}_r^{(n_{r})}\}_{n_r=1}^\infty$ of a common parameter $\theta$
whose distribution for sufficiently large $n_r$
can be approximated as
$\hat{\theta}_r^{(n_{r})} \g \theta \sim N(\theta,
(\sigma_r^{(n_r)})^2)$ with  $\sigma_r^{(n_r)}$ known.
Denote with $BF_R^{(n_r)}$ the replication BF based on $\hat{\theta}_r^{(n_{r})}$.
 Let $\theta^*$ denote the true value of $\theta$. Then the following
   limits in probability hold
\begin{align}
\label{eq:consistency-BFR}
\begin{split}
\mbox { if }  \theta^*=0, & \quad BF_R^{(n_r)}  \underset{n_r \rightarrow \infty}{\rightarrow} \infty   \mbox { at rate } {O(\sqrt{n_r})}; \\
\mbox { if }  \theta^* \ne 0, & \quad  BF_R^{(n_r)}  \underset{n_r \rightarrow \infty}{\rightarrow} 0   \mbox { at rate } {\exp\{-K n_r \}},
\end{split}
\end{align}
where
 $K>0$ is a positive constant.
 As a consequence, $BF_R$ is \emph{consistent}.

\end{proposition}

\emph{Proof}. See Supplementary material.

\begin{proposition}
\label{prop:inconsistency-BFSA}
Under the assumptions   of Proposition 1,
let $p_S(\cdot)$ and  $p_A(\cdot)$ denote the density of the skeptical and the advocate prior leading to \eqref{eq:BFSA}.
 Let $\theta^*$ denote the true value of $\theta$. Then the following limit in probability holds
\ben
\label{eq:inconsistency-BFSA}
BF_{S:A}(\hat{\theta}_r^{(n_{r})}; g) & \underset{n_r \rightarrow \infty}{\rightarrow} \frac{p_S(\theta^*)}{p_A(\theta^*)}.
\een
\end{proposition}
\emph{Proof}. See Supplementary material.

  It follows from Proposition \ref{prop:inconsistency-BFSA} that consistency does not hold for $BF_{S:A}$  because it converges to a constant irrespective of the true value $\theta^*$.
 \citet{ly-wagenmakers2022bayes}, discussing Bayes factors for \lq \lq peri-null\rq \rq{} hypotheses, also mention,  as a particular case,  the inconsistency of $BF_{S:A}$.
We note that both the consistency of $BF_R$ and the inconsistency of $BF_{S:A}$
reported in Proposition \ref{prop:consistency-BR} and
Proposition \ref{prop:inconsistency-BFSA}, respectively,
are  in accord with  theoretical results on the asymptotic  behavior  of Bayes factors under rather general conditions on model and priors presented in \citet{Dawid:2011}.

Proposition \ref{prop:inconsistency-BFSA} highlights the fact that
the pair $\{H_S; H_A \}$ leading to   $BF_{S:A}$
is a comparison between two \emph{opinions} (priors) on the parameter for the \emph{same} model.
The bottom line is that  even  an infinite replication sample size cannot favor one over the other overwhelmingly.

\subsection{Information Consistency}
\label{subsection:information consistency}

Besides consistency, another useful criterion to evaluate a Bayes factor
is \emph{information consistency}. \citet{bayarri2012criteria} present this criterion with regard to two \emph{nested} models,   $M_0$ and $M$, with $M_0$ (the null  model) nested in $M$. Let  $\Lambda_{M_0:M}(y)$ be the likelihood ratio, and  consider a sequence of data vectors $\{ y_m \}$ of \emph{fixed} sample size,
such that
\ben
\label{eq:Lambda}
\lim_{m \rightarrow \infty} \Lambda_{M:M_0}(y_m)=\infty,
\een
so that, in the limit,
the data  provide overwhelming evidence in favor of $M$. It is then required that the BF in favor of $M$ follows suit and diverges accordingly.
We show in the Supplementary material (Proposition S.1) that $BF_R$ is information consistent.
On the other hand, the noton of information consistency becomes vacuous when it comes to the skeptical $BF_{S:A}$,  because the two models under comparison are not nested, as already mentioned at the end of Section \ref{subsection:the-skeptical}.
We note that  some concerns about  information consistency are  addressed in \citet[Section 3.4]{pawel2022sceptical},
to whom we refer for further details.

\subsection{Prior-Data Conflict}
\label{subsection:prior data}

\color{black}

The skeptical prior is constructed by fixing a value $\gamma$ for $BF_{0:S}$ which renders the original finding unconvincing at level $\gamma$.
How  reasonable is the skeptical prior  $\theta \sim N(0,g_{\gamma} \sigma_o^2)$  relative to  the information on $\theta$ provided by the data in the original experiment?
\black
The same concern  applies to  $N(0,g_{S} \sigma_o^2)$, where $g_S$  is the value corresponding to $BF_S$ defined in Equation \eqref{eq:BF-S}.
Surely a skeptical prior which is at odds with the original data would appear  suspicious  to an external agent (e.g.  a regulatory agency or a project grant reviewer). Indeed we would like to be skeptical but not unrealistic.
The problem we are addressing  is an instance of \emph{prior-data-conflict} \citep{evans2006checking, egidi2021avoiding}; see also
\citet{held2020new} in the context of replication studies.

Here we simply sketch the idea.
Consider a statistic $T$ having distribution $f_T(t|\theta )$ and a prior  $\theta \sim \pi(\theta)$.
Let the marginal density be given by
\begin{equation}
m_T(t) = \int f_T(t| \theta) \pi(\theta)d\theta,
\label{eq:marginal density of T}
\end{equation}
where $t$ ranges over the set of values of $T$. Let $t_{obs}$ be the observed value of $T$. The $p$-value for prior-data conflict \citep{evans2006checking}  is defined as:
\begin{equation}
P(t_{obs})= \text{Pr}^{m_T} \{ t: m_T(t) \leq m_T(t_{obs}) \},
\label{eq:2}
\end{equation}
where $P^{m_T}(\cdot)$ is the probability computed under the marginal $m_T(\cdot)$ in \eqref{eq:marginal density of T}.
The index $P(t_{obs})$ measures how surprising is the value $t_{obs}$  by computing the probability of all those $t$'s whose density is below the density at $T=t_{obs}$. Intuitively, if $P(t_{obs})$ is very small, say  below 10\% or possibly a smaller value, then a surprising value has occurred and in this case we declare prior-data conflict. If $m_T(\cdot)$ is unimodal, $P(t_{obs})$ provides the tail probabilities under $m_T(\cdot)$, where the tails are the $t$-values whose density is below the cutoff $m_T(t_{obs})$.

\section{ The Skeptical Mixture Prior and Bayes Factor}
\label{sec:the skeptical}

We generalize the skeptical  prior $H_S: \theta \sim {N}(0, g_\gamma \sigma^2_o)$  employed in \eqref{eq:H-S} with a \emph{mixture} prior composed of a point mass  and a continuous component.
These type of priors  have been already implemented
\citep{rockova2018bayesianestimation}
as variants of the classic spike-and-slab prior;  they have also  been used as data distribution in genomic studies   \citep{TaylorPollard2009hypothesistests}.

Specifically we define the   \emph{family of skeptical mixture priors} at level $\gamma \in (0,1)$ as
\ben
\label{eq:skeptical-mixture-prior}
\theta \sim
\psi_\gamma \delta_{0} + (1-\psi_\gamma) {N}(0, h_\gamma \sigma^2_o), \quad (\psi_\gamma,h_\gamma) \in U_\gamma,
\een
 where $\delta_{0}$ is the Dirac measure at $\theta=0$,  $0 \leq \psi_\gamma  \leq 1$ is  a weight, $h_\gamma>0$ is the relative variance, and $U_\gamma$ is the set of pairs
$(\psi_\gamma, h_\gamma)$ such
that the BF for the comparison of $H_0: \theta=0$ against
the hypothesis that $\theta$ follows \emph{any} distribution   in the family \eqref{eq:skeptical-mixture-prior}
in the original experiment
is equal to $\gamma$.

 As for the skeptical prior, our notation does not make explicit the dependence of the pair
$(\psi_\gamma,h_\gamma)$
 on the original estimator $\hat{\theta}_o$.
It is worth emphasizing that, differently from the skeptical prior, its mixture  counterpart  is naturally defined as a \emph{set} of distributions.

The family of priors in \eqref{eq:skeptical-mixture-prior}  includes the skeptical prior \eqref{eq:H-S} as a special case by setting $(\psi_\gamma=0, h_\gamma=g_\gamma)$.
The family is clearly empty if
the condition that the BF is equal to $\gamma$ cannot be fulfilled.

\subsection{Prior-data conflict under the skeptical mixture prior}
\label{subsec:pd_mix}

Consider $\hthetao|\theta \sim N(\theta, \sigma^2_o)$, with $\sigma^2_o$ known, and assume
that $\theta$ is distributed according to the {skeptical mixture prior} at level $\gamma$, \eqref{eq:skeptical-mixture-prior}.
The marginal density of the estimator $\hthetao$ is
$$m(\hthetao) = \int N(\hthetao|\theta, \sigma^2_o) dF_{SM}(\theta),$$
where  $N(\hthetao|\theta, \sigma^2_o)$ is a shorthand notation for the sampling density of $\hthetao$ and $F_{SM}(\theta)$
the cdf of the mixture prior \eqref{eq:skeptical-mixture-prior}.
We obtain
\begin{equation}
m(\hthetao) = \psi_\gamma N(\hthetao|0, \sigma^2_o) + (1-\psi_\gamma) N(\hthetao|0, \sigma^2_o(1+h_\gamma)).
\label{eq:mhthetao}
\end{equation}

Simplifying the notation,
 the structure of Equation \eqref{eq:mhthetao}
 can be formally written  as
\begin{equation}
m_T(t) = \psi N(t|0, \sigma^2) + (1-\psi) N (t|0, \sigma^2(1+h)).
\label{eq:1}
\end{equation}

\black
To evaluate the $p$-value for prior-data conflict $P(t_{obs})$
defined in \eqref{eq:2}
with regard to  \eqref{eq:1}
it is expedient
to introduce an auxiliary  random variable $V$ having a  $\text{Bern}(\psi)$ distribution and define the joint density of $(T,V)$ as $h(t,v|\theta)=f(t|v,\theta)g(v)$ where
$g(0)=\psi$,   $g(1)=(1-\psi)$ and
\begin{equation*}
f(t|v,\theta) =
\begin{cases}
N(t|0,\sigma^2) &\textrm{ if} \quad v=0\\
 N(t|\theta,\sigma^2) &\textrm{ if} \quad  v=1.
\end{cases}
\end{equation*}
Let $\theta \sim N(\theta|0,\sigma^2 \cdot h)$. Then, marginally
\begin{align*}
h(t)&= \sum_v \left\{  \int h(t,v |\theta) N(\theta|0,\sigma^2 \cdot h) d\theta \right\}g(v)\\
&=\psi \int N(t|0,\sigma^2)p(\theta)d\theta+
(1-\psi) \int N(t|\theta,\sigma^2)p(\theta)d\theta\\
&=\psi  N(t;0,\sigma^2)+(1-\psi)  N(t;0,\sigma^2 \cdot (1+h)),
\end{align*}
which coincides with \eqref{eq:1}.

Since $V$ is ancillary, one can condition on it to compute prior-data conflict; see \citet{evans2006checking}.
Hence
\begin{align*}
\text{P}(t_{obs}|v=0)&=\text{Pr} \left\{ N(T|0,\sigma^2) \leq N(t_{obs}|0,\sigma^2)  \right\}\\
\text{P}(t_{obs}|v=1)&=\text{Pr} \left\{ N(T|0,\sigma^2(1+h)) \leq N(t_{obs}|0,\sigma^2)(1+h)  \right\},
\end{align*}
whence
\begin{equation}
\label{eq:Ptobs}
\text{P}(t_{obs})=\psi \text{P}(t_{obs}|v=0)+(1-\psi)\text{P}(t_{obs}|v=1).
\end{equation}

\black

\black

\begin{lemma*}
Let $T \sim f(t) = N(t|0, \tau^2)$. Then
\begin{equation*}
\text{Pr} \{f(T) \leq f(t_{obs})  \}= \text{Pr} \left\{ U \geq \left( \frac{t_{obs}}{\tau} \right)^2 \right\},
\end{equation*}
where $U \sim \chi^2(1)$, a chi-squared distribution with one df.
\end{lemma*}

\emph{Proof}: $f(T) \leq f(t_{obs})$ iff $ \left( \frac{T}{\tau} \right)^2 \geq  \left( \frac{t_{obs}}{\tau} \right)^2$, and 
$(T/\tau)^2 \equiv U \sim \chi^2(1)$. $\Box$

Using the lemma together with \eqref{eq:Ptobs}  and reverting to the notation used in
\eqref{eq:mhthetao},
  the $p$-value for prior-data conflict based on the skeptical mixture prior \eqref{eq:skeptical-mixture-prior} is
\begin{equation}
P_\gamma(\hthetao)= \psi_\gamma (1-G_1(z^2_o)) + (1-\psi_\gamma)(1-G_1(z^2_o/(1+h_\gamma))),
\label{eq:p_value}
\end{equation}
where $G_1(\cdot)$ is the cdf of a chi-squared distribution with one df.

\subsection{Embedding control for prior-data conflict in the skeptical mixture prior}
\label{sec:embedding}

Since any element in the \emph{set} $U_\gamma$ of hyperparameters
$\{(\psi_\gamma, h_\gamma)  \}$
describing the family \eqref{eq:skeptical-mixture-prior}
leads to a BF equal to $\gamma$,
a skeptic is offered the opportunity to select  a prior which in addition
 provides some control on prior-data conflict.

\begin{figure}[H]
\begin{center}
\includegraphics[scale=0.5]{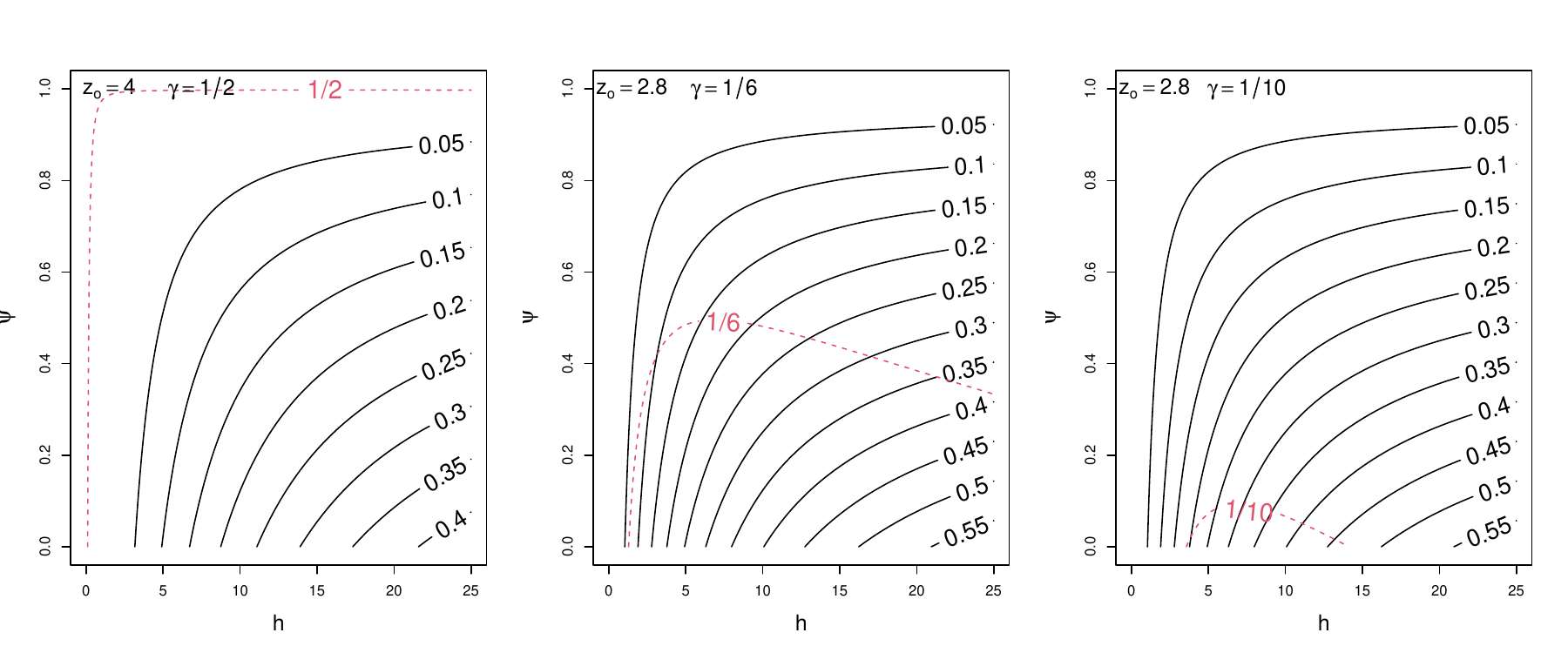}
\end{center}
\caption{
 Contours of $p$-values  for prior-data conflict (solid black line).
 Contour for $BF_{0:SM}(\hat{\theta}_o; \psi,h)= \gamma$ (dashed  red line), for three selected values of $z_o$ and $\gamma$. 
\label{fig:first}}
\end{figure}

Based on the analysis reported in Section \ref{subsec:pd_mix} and in particular Equation
\eqref{eq:p_value},
  let
\begin{equation}
(\psi_{\gamma,\alpha}, h_{\gamma,\alpha}) \in U_\gamma \,: \,
P_\gamma(\hthetao) =\alpha,
\label{eq: alpha_condition}
\end{equation}
be a pair
defining a skeptical mixture prior in
\eqref{eq:skeptical-mixture-prior}
whose $p$-value for prior-data conflict,
$P_\gamma(\hthetao)$,  is equal to $\alpha$. Small values of $\alpha$ indicate that the observed value $\hthetao$ is highly unlikely to occur under the skeptical mixture prior.  One could set for instance $\alpha=0.1$ as an upper-bound for the $p$-value  to declare  \emph{lack} of prior-data conflict.
\black
The problem of finding  $(\psi_{\gamma,\alpha}, h_{\gamma,\alpha})$ can be visually represented by plotting,
in  $(h,\psi)$  space:
i) the contour lines realizing $P(\hthetao; \psi, h)=\alpha$ for a grid of $\alpha$-values, where
$P(\hthetao; \psi, h)$ is identical to the expression in \eqref{eq:p_value}
except that $(h,\psi)$ are \emph{unconstrained}:
ii)
the contour line  $U_\gamma$ for a fixed value of $\gamma$;
  and finally looking for possible points of intersection.
 Figure \ref{fig:first} illustrates this procedure with varying $z_o$ and $\gamma$.  \color{black}
 We remark that the general shape of the contours $P(\hthetao; \psi, h)=\alpha$  can be easily checked analytically
because  $\partial P(\hthetao; \psi, h)/\partial \psi<0$ for all $h>0$ and $\partial P(\hthetao; \psi, h)/\partial h>0$ for all $ 0<\psi<1$.
\black
Notice that  $P_\gamma(\hthetao) =\alpha$ might  not be attainable at the chosen level $\alpha$.
This can  happen for a variety of reasons which are depicted in Figure \ref{fig:first} for three selected values of $\gamma$.
 The first one is that an intersection is available, but it occurs   only at values $\alpha$ \emph{greater} than the predetermined threshold  (third panel in Figure \ref{fig:first})  This means that there is no appreciable prior-data conflict even for the standard skeptical prior ($\psi=0$), which can therefore be safely employed by default.
Yet another possibility is that an intersection does not occur even for very small $\alpha$-values: this goes in the opposite direction
of the case discussed above; see
  the \black first panel in Figure \ref{fig:first}\color{black}.
  This  situation will actually occur   only for few real studies  that we examine in
  Section \ref{sec:application}.
  In this case one might  settle down again  on the standard skeptical prior,
   although  one should be aware that prior-data conflict is not under control.
The final possibility is that our predetermined level $\alpha$ can indeed be achieved but not with the standard skeptical prior; this is clearly the most interesting case from our perspective and it happens in the middle panel of Figure \ref{fig:first}. In this case the standard skeptical barely achieves $\alpha=0.05$ while a skeptical mixture with a positive $\psi$ and correspondingly higher relative variance  $h$ achieves the goal of
$P_\gamma(\hthetao)=0.1$

Consider now the comparison
    \ben
    \label{eq:SM-versus-A}
    H_{SM}: \theta \sim \psi_{\gamma,\alpha} \delta_{0} + (1-\psi_{\gamma,\alpha}) {N}(0, h_{\gamma, \alpha} \sigma^2_o)
    \quad   \mbox{versus} \quad
     H_A: \theta \sim {N}(\hat{\theta}_o, \sigma^2_o),
\een
where $H_{SM}$ represents the skeptical mixture prior and $H_A$ the advocate prior.
Let $p_A(\cdot)$ be the density function of the advocate prior and let
$f(\hat{\theta}_r \g H_A)=\int f(\hat{\theta}_r \g \theta)p_{A}(\theta) d\theta$ denote the marginal density of $\hat{\theta}_r$ conditionally on $H_A$.
Similarly let
$p_S(\theta; h_{\gamma, \alpha})= N({\theta}; 0,h_{\gamma, \alpha} \sigma_o^2 )$ be the density function of the continuous component of the  skeptical mixture prior and let
 $f(\hat{\theta}_r \g H_{S}, h_{\gamma, \alpha})=\int f(\hat{\theta}_r \g \theta)
 p_{S}(\theta; h_{\gamma, \alpha} ) d\theta$ denote the marginal density of $\hat{\theta}_r$ conditionally on $H_{S}$ with given $h_{\gamma, \alpha}$, as in \eqref{eq:H-S}.
Finally, let   $P_{SM}(\cdot)$ be the cdf of the skeptical mixture prior
\eqref{eq:SM-versus-A},
 which is everywhere continuous,  save in $\theta=0$, where it makes a jump equal to $\psi_{\gamma,\alpha}$.
Then
\begin{align}
BF_{SM:A}(\hat{\theta}_r; \psi_{\gamma, \alpha}, h_{\gamma, \alpha})
=
&\
\frac{\int f(\hat{\theta}_r \g \theta)dP_{SM}(\theta)}
{f(\hat{\theta}_r \g H_A)} \nonumber \\
=
&\
\frac{1}{f(\hat{\theta}_r \g H_A)}
 \times
 \left(
 \psi_{\gamma, \alpha}
{f(\hat{\theta}_r \g \theta=0)}
 + (1-\psi_{\gamma, \alpha}) f(\hat{\theta}_r \g
 H_S)  
 \right) \nonumber \\
 = &\ \psi_{\gamma, \alpha} BF_R + (1-\psi_{\gamma, \alpha})
BF_{S:A}(\hat{\theta}_r; h_{\gamma, \alpha}),
\label{eq:mixture_bf}
\end{align}
where
$BF_R$ is the replication BF defined in \eqref{eq:BR}, and $BF_{S:A}$ is the  BF comparing the skeptical and the advocate prior defined in \eqref{eq:BFSA} having relative variance
$h_{\gamma, \alpha}$,
which now depends  on $\gamma$ as well as $\alpha$.
We highlight that $BF_{S:A}$ appearing in \eqref{eq:mixture_bf} is not the standard skeptical BF of \eqref{eq:BFSA} because it is evaluated at the relative variance $h_{\gamma, \alpha}$, and thus already incorporates the prior-data conflict constraint.
We then declare \emph{replication success} at level $\gamma$ iff
\begin{align*}
BF_{SM:A}(\hat{\theta}_r; \psi_{\gamma, \alpha}, h_{\gamma, \alpha}) \leq \gamma,
\end{align*}
that is, the data favor the advocate over the sceptical mixture prior at a higher level than the skeptic's initial objection.
Analogously to the skeptical Bayes factor of \cite{pawel2022sceptical} in Equation~\eqref{eq:BF-S},     the \emph{skeptical mixture Bayes factor} is defined as
\begin{equation}
{BF}_{SM}(\alpha) = \inf \{\gamma : {BF}_{SM:A}(\hat{\theta}_r; \psi_{\gamma, \alpha}, h_{\gamma, \alpha}) \leq \gamma \},
\label{eq:BF-SM}
\end{equation}
with $\alpha$  a further tuning parameter.
As for the skeptical Bayes factor it may happen that  $BF_{SM}(\alpha)$ does not exist, because there is no $\gamma$ for which replication success can be established, but this usually occurs when $|z_o|$, or $|d|=|z_r|/|z_o|$,  or both are too small.

The following represents an important feature of our proposal.

\begin{result}
Under the skeptical mixture prior introduced in
\eqref{eq:SM-versus-A}, if
 $\psi_{\gamma, \alpha}>0$ and the true value is $\theta^*=0$,
 then
$BF_{SM:A}(\hat{\theta}_r; \psi_{\gamma, \alpha}, h_{\gamma, \alpha})$ is consistent.
\end{result}
\emph{Proof}.
For  $n_r \rightarrow \infty$,
the result follows immediately from \eqref{eq:mixture_bf} and the fact that
 $BF_R \rightarrow \infty$,  if $\theta^*=0$     because
of \eqref{eq:consistency-BFR},
while $BF_{S:A}(\hat{\theta}_r; h_{\gamma, \alpha})$
converges to a constant because of \eqref{eq:inconsistency-BFSA}. $\Box$

Thus,  if the effect is truly absent,  this will be  flagged by $BF_{SM:A}$ with unlimited evidence if the sample size grows indefinitely. On the other hand if $\theta^* \neq 0$, then the continuous skeptical component of the mixture will take the lead,  and $BF_{SM:A}$ will converge to the constant
$(1-\psi_{\gamma, \alpha}) \frac{p_{S}(\theta^*; h_{\gamma,\alpha})}{p_A(\theta^*)}$; see Proposition \ref{prop:inconsistency-BFSA}.
While this result is only partial,   it is particularly  useful in a replication setting wherein correctly ascertaining the lack of an effect may prove very valuable to contrast an original finding possibly pointing in a different direction.

To better appreciate the skeptical mixture BF, we further investigate the behavior of $BF_{SM}(\alpha)$ in  \eqref{eq:BF-SM}.
The following preliminary fact about the skeptical BF  will be useful in the sequel.
Recall that the value $BF_{0:S}(\hat{\theta}_o;  h_\gamma)=\gamma$, when it exists,  can be reached for two values of $h_\gamma$, namely $h_\gamma=g_\gamma$ and $h_\gamma=g_\gamma^{JL}$, where $g_\gamma<g_\gamma^{JL}$. The former is the classic skeptical relative variance of \citet{pawel2022sceptical}; while the latter, corresponding to a higher variance, arises because of the Jeffreys-Lindley's paradox described for instance in
\citet[sect. 6.1.4]{Bernardo:Smith:2000}
and illustrated in \citet[Fig. 2]{pawel2022sceptical}.

For fixed $\hat{\theta}_{o ,obs}\neq 0$ and $(\psi,h) \in \Re^+ \times \Re^+$
the partial derivative of $P_\gamma(\hat{\theta}_{o ,obs}, \psi,h)$ wrt $\psi$ is negative, while that wrt $h$ is positive; see also Figure 1.
The reason why this occurs is because increasing $\psi$  while holding $h$ fixed will subtract area from the tails of
$m(\hthetao, \psi,h)$ where the density is below $m(\hat{\theta}_{o,obs}, \psi,h)$. Thus, to increase $\alpha$ one should let $\psi$ diminish and $h$ increase;  see also Figure 1 in the Supplementary material for a graphical explanation. \black
The locus of points $U_\gamma$ traverses the contours of
 $P_\gamma(\hthetao, \psi,h)$, see Figure \ref{fig:first}, and as $h$ increases along $U_\gamma$, $\alpha$ also increases, while $\psi$ eventually will decrease to zero.
 Suppose now the supremum of $\alpha$ having an intersection with $U_\gamma$ exists, and let it be
 $\alpha_\gamma^*$.
 Then as $\alpha \rightarrow \alpha_\gamma^*$
 $$
 \psi_{\gamma, \alpha} \rightarrow 0; \quad  h_{\gamma, \alpha} \rightarrow g_\gamma^{JL}.
 $$
 The above result holds for each $\gamma$. Because of \eqref{eq:BF-SM} we can thus conclude that as $\alpha \rightarrow \alpha_\gamma^*$
$$
 BF_{SM}(\alpha) \rightarrow BF_S,
$$
where $BF_S$ is the skeptical Bayes factor defined in \eqref{eq:BF-S}. The result holds because
$BF_{S:A}(\hat{\theta}_r; g_{\gamma})=BF_{S:A}(\hat{\theta}_r; g_{\gamma}^{JL})$.

\subsection{Example}
\label{subsec:examples}

To illustrate our method and provide a comparison we use   the same setting  discussed in \citet[Sect. 2.2]{pawel2022sceptical}, and accordingly fix   \black $z_o=3, \ z_r =2.5$ and
$c= \sigma^2_r/\sigma^2_o=1$,  so that
$d=\hat{\theta}_r/\hat{\theta}_o = 0.83$.
This setup is meant to represent a situation often encountered in practice  with the  replication study providing a somewhat weaker evidence against the null  than the original study.
Additionally we fix the  $p$-value for prior-data conflict at level $\alpha = 0.1$.

\begin{figure}[H]
\begin{center}
\includegraphics[scale=0.9]{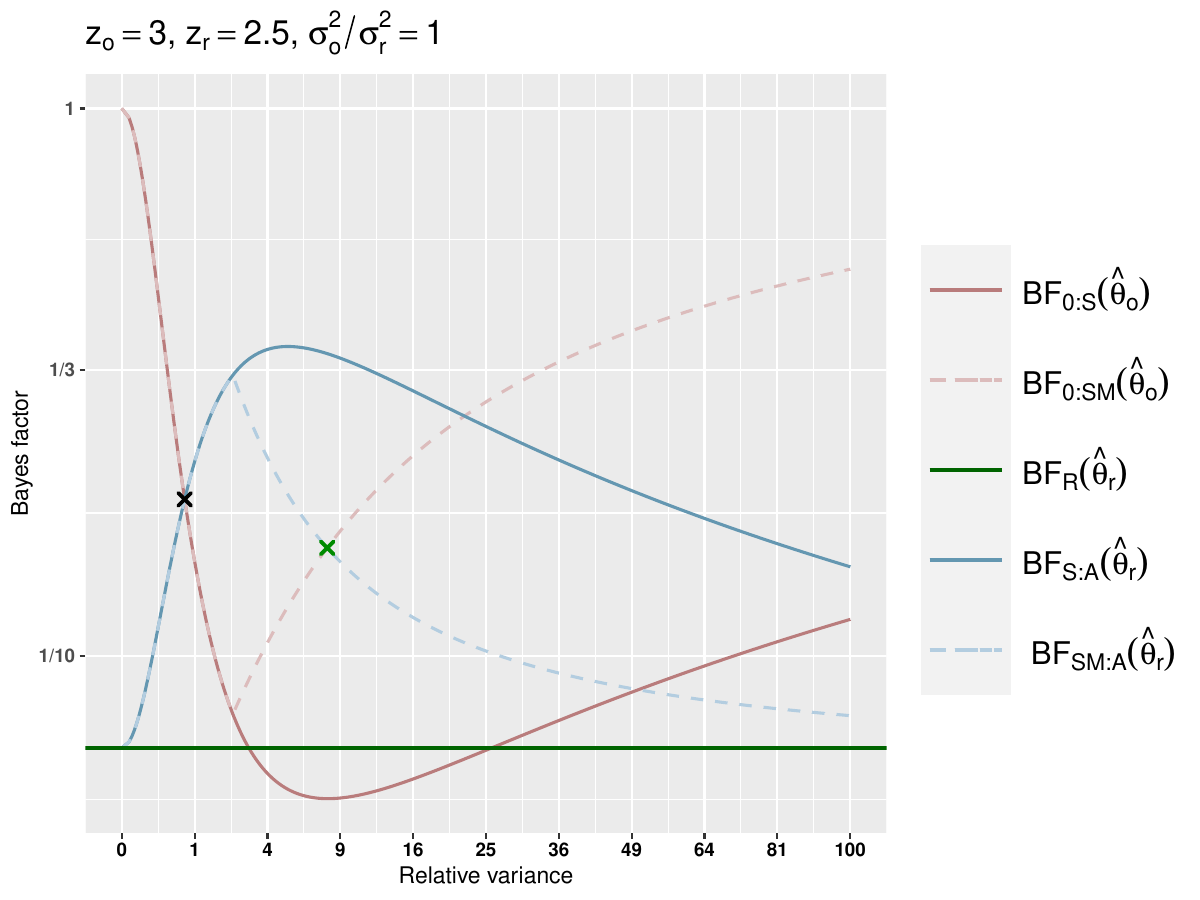}
\end{center}
\caption{Bayes factors $BF_{0:S}(\hat{\theta}_o;g)$, $BF_{S:A}(\hat{\theta}_r; g)$, $BF_{0:SM}(\hat{\theta}_o; \psi, h)$, $BF_{SM:A}(\hat{\theta}_r; \psi, h)$ and $BF_R(\hthetar)$ as a function of the relative variance.
  The black cross represents the skeptical BF, $BF_{S}$,  while the green one
 represents the skeptical mixture BF, $BF_{SM}(\alpha)$,
 with  $(\psi_{\gamma, \alpha}, h_{\gamma, \alpha})$
 evaluated at $\gamma_{SM}= BF_{SM}(\alpha)$ and $\alpha=0.1$ under the original data.}
\label{fig:second}
\end{figure}

\begin{center}
INCLUDE HERE Figure \ref{fig:second}
\end{center}

We compute the skeptical and the skeptical mixture BF proposed in Equations~\eqref{eq:BF-S}, \eqref{eq:BF-SM}, equal to 0.19 and 0.16, respectively.
In this case the skeptical relative variance can be shown to be  $g_\gamma=0.75$.
Moreover,  $(\psi_{\gamma, \alpha=0.1}, h_{\gamma, \alpha=0.1})=(0.69, 8.16)$: see also Figure 2 in the Supplementary material for a graphical explanation.
Turning to Figure \ref{fig:second} we see four curves. Two, namely  $BF_{0:S}$  (solid dark brown) \black
and $BF_{0:SM}$  (dashed light brown), \black  are based on the original data, while
 $BF_{S:A}$  (solid dark blue) \black and $BF_{SM:A}$  (dashed light blue), \black  refer to replication data.
 When two curves are basically superimposed they appear as dashed   with alternating dark and light color. \black
  The black cross represents the skeptical BF, $BF_{S}$,  while the green one  represents the skeptical mixture BF, $BF_{SM}(\alpha)$. \color{black}  All curves are plotted as a function of the relative variance.
 Additionally all skeptical mixture priors   realize a $p$-value for prior-data conflict equal to $\alpha=0.1$.
 The replication Bayes factor $BF_R$ is also included, and  appears  as a constant  green line because its corresponding prior has no hyperparameters.
 Notice that $BF_{0:S}(\hthetao)$ and $BF_{0:SM}(\hthetao)$ initially decrease, and essentially
 coincide, up to a certain level of the relative variance.
  After this threshold the
  $BF_{0:SM}(\hthetao)$ curve   starts
increasing whereas
   $BF_{0:S}(\hthetao)$  continues to decrease for a while and then  starts increasing, too.
\black
This behavior can be explained as follows.
Recall that  $z_o=3$ is a result which exhibits evidence against the null.
Consider the skeptical prior first.
As the relative variance increases it will push  mass towards  areas in the $\theta$-space better supported by the data, and so evidence for the null
will initially decrease; then  it will make a turn and start increasing (in the interval between 4 and 9): this   happens because as the variance increases further  mass is pushed away  in the tails of the $\theta$-space; the end result is   that $H_0$ becomes more reasonable (Jeffreys'-Lindley paradox).
Now turn to the curve under the skeptical mixture prior. In the first part it  essentially coincides with that for the skeptical prior (monotone decreasing behavior),  but then it  makes a bend and starts increasing at an earlier stage than the skeptical prior curve  \black (this is about half-way between 1 and 4 on the horizontal axis).
The reason why this occurs
is because the skeptical mixture   incorporates a constraint on prior-data conflict which is absent in the skeptical prior.
Specifically,  pairs $(h,\psi)$  on the same  prior-data conflict contour level are positively related; see
Figure \ref{fig:first}.
 This implies that, as the relative variance increases so does $\psi$; in this way  the evidence for $H_0$ is further  enhanced because it can benefit from two sources: the increased  variance and the greater lump mass on $\theta=0$ in the prior. This explains why the $BF_{0:SM}(\hthetao)$ curve starts turning upward at an earlier stage.
\black

A mirror-like phenomenon  occurs for the
curves $BF_{S:A}(\hthetar)$ and $BF_{SM:A}(\hthetar)$ because $z_r=2.5$,  so that evidence is still against the null, although to a lesser extent.  Now the curves will first increase
together because mass is pulled away from areas around zero which are not supported by the likelihood, and will thus favor the skeptical, respectively skeptical mixture,  hypothesis (recall that they are both centered on $\theta=0$.
Both curves will then start decreasing
basically for the same reason explained earlier, namely that as the variance increases more prior mass escapes to areas of the parameter space hardly supported  by the likelihood and this makes a fixed prior, like the advocate prior,  more reasonable.
Again change in monotonicity happens at an earlier stage for the   skeptical mixture prior,  because it must control for prior-data conflict.
Another feature to be noted is that
$BF_{SM:A}$ lies between the upper curve  $BF_{S:A}$ and the lower one $BF_R$ coherently with
Equation \eqref{eq:mixture_bf}.

\begin{figure}[H]
\begin{center}
\includegraphics[scale=0.9]{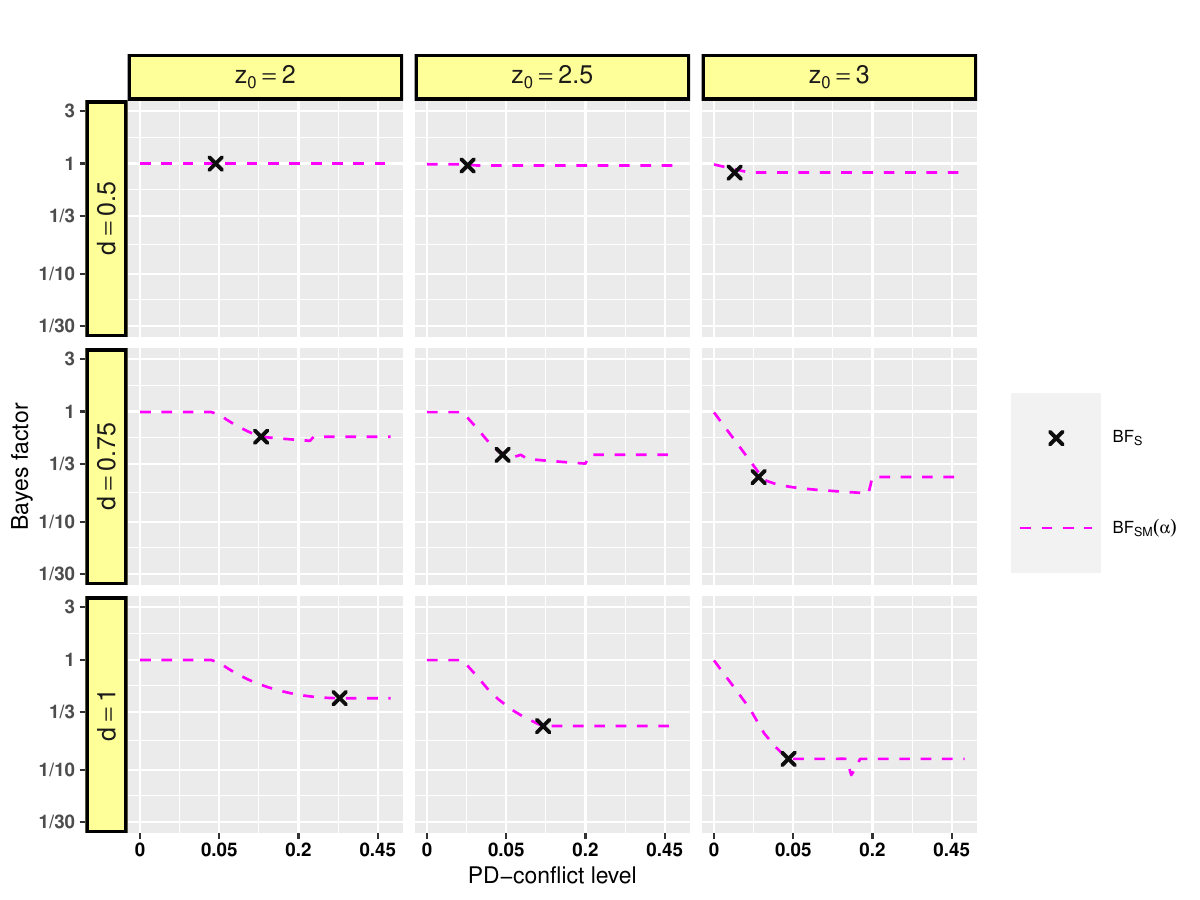}
\end{center}
\caption{Skeptical and skeptical mixture Bayes factors $BF_S$ and $BF_{SM}(\alpha)$ for varying $z_o$ and $d$ as functions of the prior-data conflict threshold $\alpha$. In all examples $c = \sigma^2_o/\sigma^2_r = 1$.}
\label{fig:third}
\end{figure}

Finally, we assess the behavior of $BF_{SM}(\alpha)$ as a function of $\alpha$,
in order to highlight the role of the prior-data conflict threshold in terms of  replication success. To this end we plot some scenarios in Figure \ref{fig:third},
\black
 where  $c=1$, $z_o \in \{2, 2.5, 3\}$ and the ratio $d= \hthetar/\hthetao \in \{1,  0.75,0.5 \}$.
The skeptical Bayes factor is marked by a black cross in correspondence of the realized level of prior-data conflict attained by the skeptical prior of \cite{pawel2022sceptical}, whereas the skeptical mixture Bayes factor varying with $\alpha$ is denoted by a pink dashed line.
As we argued at the end of Section 2.2, one can see that the skeptical mixture BF stabilizes around the  $BF_S$ as $\alpha$ grows.
Also it appears that the  prior-data conflict realized under the skeptical prior increases ($\alpha$ decreases) as $z_o$ increases.
In particular for $z_o=3$ the value of $\alpha$ is always below 5\% suggesting incompatibility of the skeptical prior with the data in the original study.

\begin{center}
INCLUDE HERE Figure \ref{fig:third}.
\end{center}

\section{Case studies}
\label{sec:application}

In this section we consider  real data sets 
  from the \emph{Social Sciences Replication Project} (SSRP), \citep{camerer2018evaluating}.
 In 2016 SSRP planned to replicate a collection of experimental studies in the social sciences published in \emph{Nature} and \emph{Science} in the period  2010-2015. These studies were  chosen because they were published in  two high-profile journals,  share a common experimental  structure and   test a treatment effect with a statistically significant finding.
In particular, we compare the results obtained  from 12 all without missing data using our skeptical mixture prior methodology with those
  produced by the skeptical BF as well as the replication BF.
Effect estimates for each study
were reported on the correlation scale $r$.
Following \citet[Section 5 and Table 2]{pawel2022sceptical},
Fisher $z$-transformation was applied to obtain approximate normality
 for the estimator $\hat{\theta}=tanh^{-1}(r)$,
and moreover $c \approx n_r/n_o$.
Throughout  we evaluated the skeptical BF, $BF_S$, and the skeptical mixture BF, $BF_{SM}(\alpha)$,
along with the target  $p$-value for prior-data conflict for three selected thresholds, $\alpha= \{ 0.01, 0.05, 0.1 \}$ when applicable.

\begin{figure}[H]
\begin{center}
\includegraphics[scale=0.74]{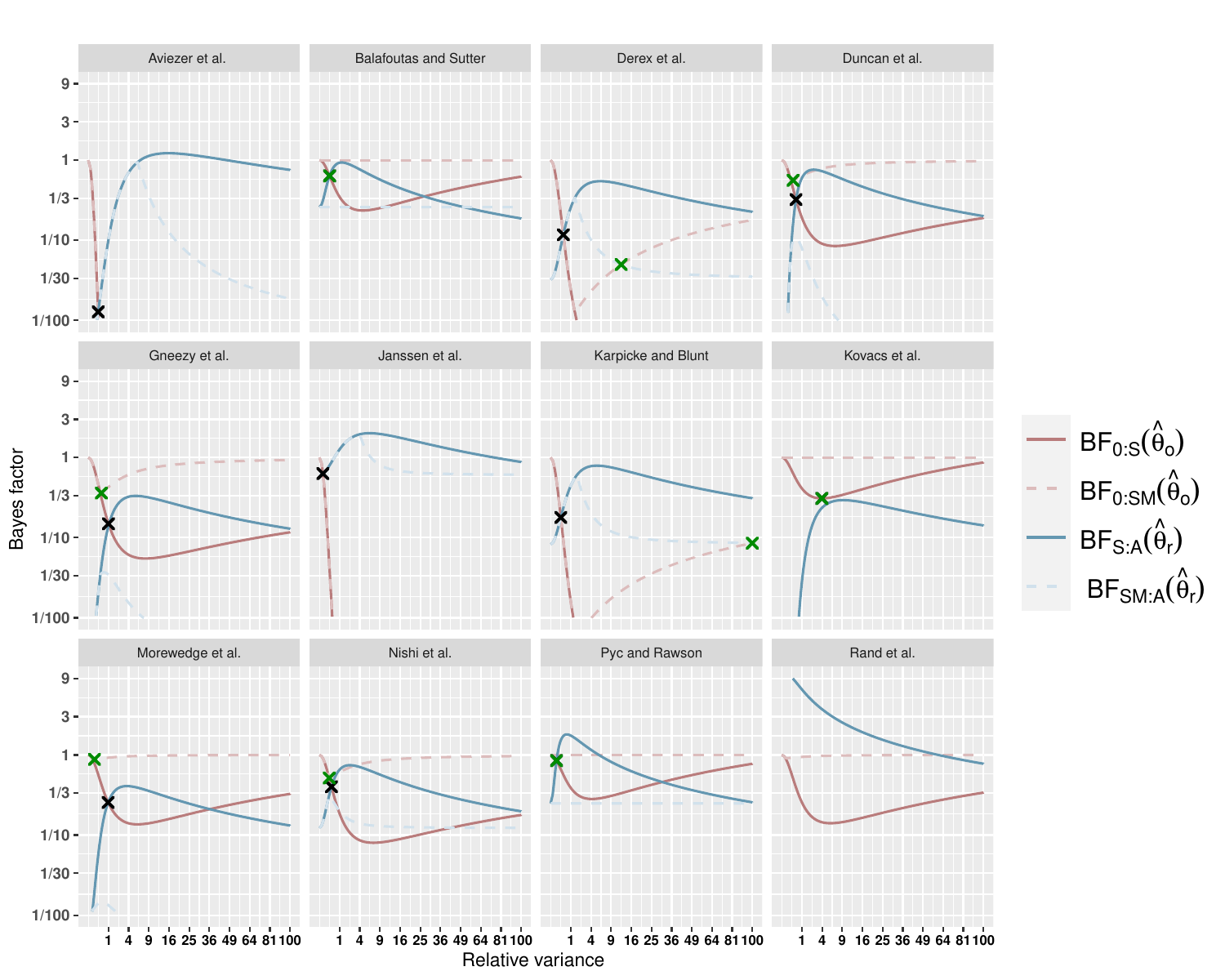}
\caption{Bayes factors $BF_{0:S}(\hat{\theta}_o;g)$, $BF_{S:A}(\hat{\theta}_r; g)$, $BF_{0:SM}(\hat{\theta}_o; \psi, h)$, $BF_{SM:A}(\hat{\theta}_r;\psi, h)$ as a function of the relative variance. Data  from twelve studies of the \emph{Social Sciences Replication Project} \citep{camerer2018evaluating}.
   The black cross   represents  the skeptical BF, $BF_{S}$,
    while the green cross represents the skeptical mixture BF,
    $BF_{SM}$, with $g_\gamma$ and $(\psi_{\gamma, \alpha}, h_{\gamma, \alpha})$ evaluated at $\gamma_S=BF_S$ and $\gamma_{SM}=BF_{SM}(\alpha)$, respectively, and $\alpha=0.01$ under the original data.
\label{fig:fourth}}
\end{center}
\end{figure}

\begin{figure}[H]
\begin{center}
\includegraphics[scale=0.74]{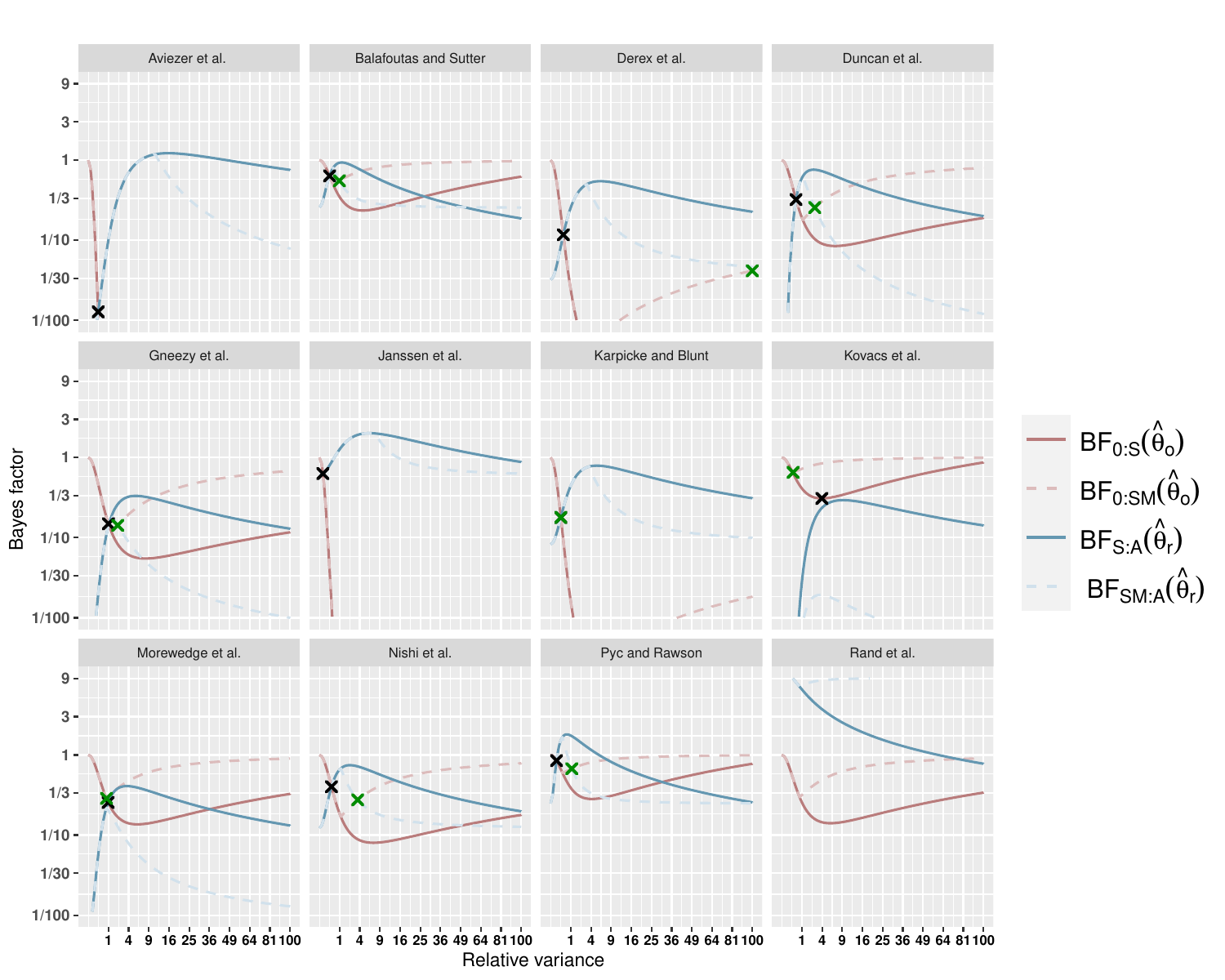}
\caption{Bayes factors $BF_{0:S}(\hat{\theta}_o;g)$, $BF_{S:A}(\hat{\theta}_r; g)$, $BF_{0:SM}(\hat{\theta}_o; \psi, h)$, $BF_{SM:A}(\hat{\theta}_r;\psi, h)$ as a function of the relative variance.  Data from twelve studies  of the \emph{Social Sciences Replication Project} \citep{camerer2018evaluating}.
   The black cross   represents  the skeptical BF, $BF_{S}$,
    while the green cross represents the skeptical mixture BF,
    $BF_{SM}$, with $g_\gamma$ and $(\psi_{\gamma, \alpha}, h_{\gamma, \alpha})$ evaluated at $\gamma_S=BF_S$ and $\gamma_{SM}=BF_{SM}(\alpha)$, respectively, and $\alpha=0.05$ under the original data.
\label{fig:fifth}}
\end{center}
\end{figure}

\begin{figure}[H]
\begin{center}
\includegraphics[scale=0.74]{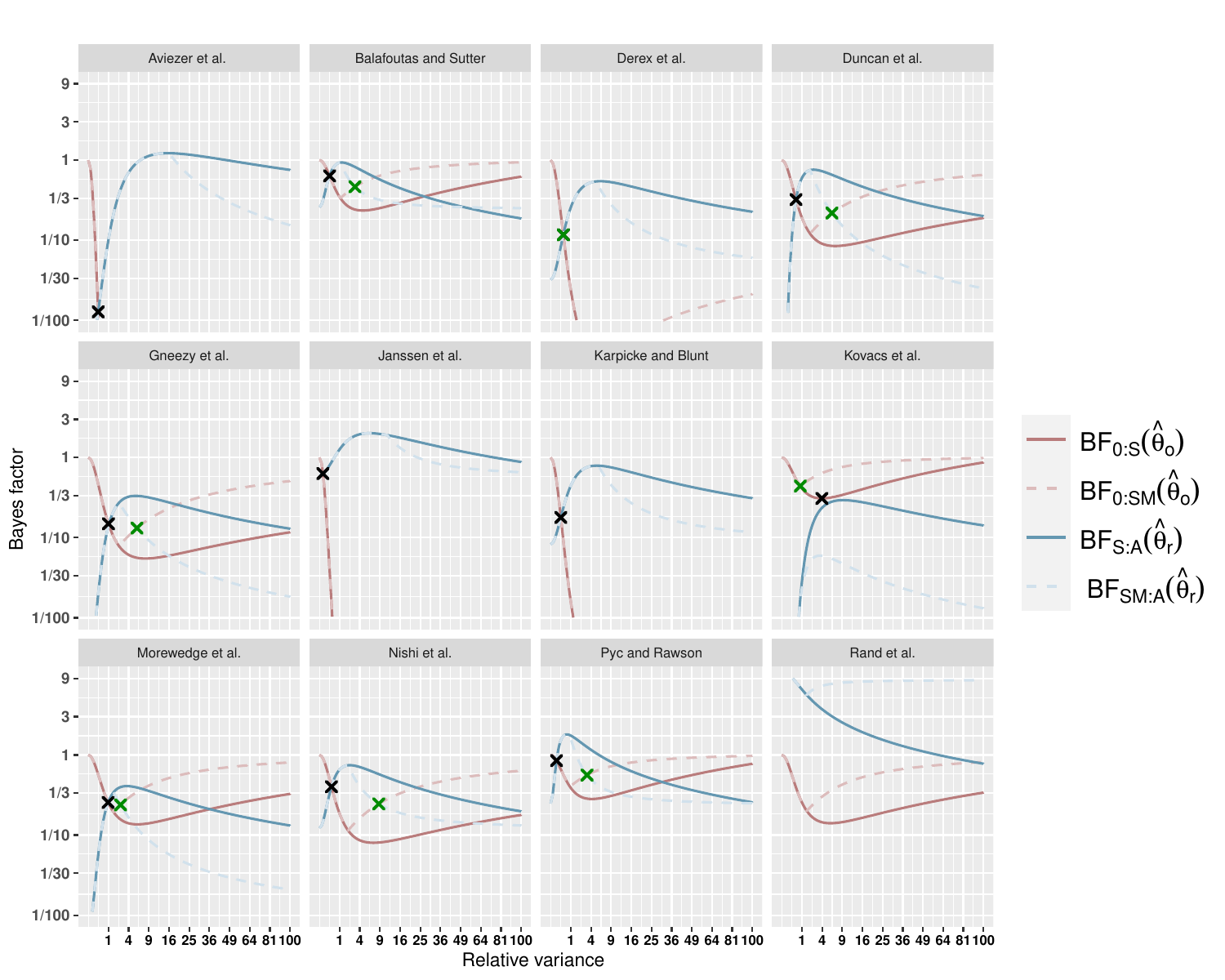}
\caption{Bayes factors $BF_{0:S}(\hat{\theta}_o;g)$, $BF_{S:A}(\hat{\theta}_r; g)$, $BF_{0:SM}(\hat{\theta}_o; \psi, h)$, $BF_{SM:A}(\hat{\theta}_r;\psi, h)$ as a function of the relative variance. Data  from twelve  of the \emph{Social Sciences Replication Project} \citep{camerer2018evaluating}.
   The black cross   represents  the skeptical BF, $BF_{S}$,
    while the green cross represents the skeptical mixture BF,
    $BF_{SM}$, with $g_\gamma$ and $(\psi_{\gamma, \alpha}, h_{\gamma, \alpha})$ evaluated at $\gamma_S=BF_S$ and $\gamma_{SM}=BF_{SM}(\alpha)$, respectively, and $\alpha=0.1$ under the original data.
\label{fig:sixth}}
\end{center}
\end{figure}

For each study, the summary statistics $\{z_o,z_r,n_o,n_r,c,d \}$ are reported  in columns 1 through 6 of Table \ref{tab:tabone} together with the hyperparameters for the skeptical priors and the three Bayes factors under investigation.
Specifically, we denote with $g_S$ and $(\psi_{SM, \alpha}, h_{SM, \alpha})$ the hyperparameters of the skeptical, respectively skeptical mixture prior,  computed in correspondence of the degrees of skepticism $\gamma_S = BF_S$ and $\gamma_{SM} = BF_{SM}(\alpha)$; see Equations  \eqref{eq:BF-S} and \eqref{eq:BF-SM}. Finally, $P_S$ and $P_{SM}$ denote the realized $p$-values---see Equation \eqref{eq:p_value}---under the skeptical and the skeptical mixture prior. The distinct scenarios reported in  Table \ref{tab:tabone}---one for each $\alpha$---are depicted in Figures \ref{fig:fourth}, \ref{fig:fifth}, and \ref{fig:sixth}.

We now summarize the main  features which emerge.

First of all we notice that the ratio $d=\hthetar/\hthetao$ is always below one, save for the study \cite{kovacs2010social}, so that the effect is less pronounced in the replication study.

Consider first the section of Table \ref{tab:tabone} with $\alpha=0.01$. Recall that, notwithstanding this value of $\alpha$, for some studies the corresponding  prior-data conflict $p$-value $P_{SM}$ may differ from this target because of the reasons we explained in Section \ref{sec:embedding}. For a few studies, notably \cite{aviezer2012body}, \cite{balafoutas2012affirmative}, \cite{janssen2010lab}, \cite{kovacs2010social}, \cite{nishi2015inequality}, and \cite{pyc2010testing}, the resulting $BF_S$ and $BF_{SM}$ are either the same or they belong to the same broad evidence  class as for instance reported in  \citet[Table 1]{schonbrodt2018bayes}. This happens because the \textit{realized} prior-data conflict is similar in the two approaches.
Turning now to  studies for which there appears to be a difference between the skeptical mixture $BF_{SM}$ and the standard skeptical $BF_S$,  an interesting pattern comes to the surface.
Start with the study \cite{derex2013experimental}, for which $P_S=0.001$ is much smaller than our set value $\alpha=0.01$. One can see that $BF_S=0.12>BF_{SM}=0.05$ (we approximate values to the second digit for ease of legibility), so that there is \textit{weaker} evidence of replication success under the skeptical prior than under  the skeptical mixture.  Interestingly this difference is meaningful on the  BF scale, because $BF_S$ corresponds to \textit{moderate} evidence in favor of  replication success, while evidence becomes \textit{strong}      when $BF_{SM}$ is considered.
The skeptical $BF_S$ achieves only moderate evidence for replication success because the resulting skeptical prior is highly conflicting with the original data ($P_S =0.001$).   Our approach   reduces the conflict to $0.01$ by making the prior variance much higher,
and this boosts the advocate prior and hence replication success.
A similar phenomenon happens with the study \cite{karpicke2011retrieval}. Conversely, consider study \cite{gneezy2014avoiding};  in this case
$P_S=0.034>P_{SM}=0.01$ and  so the resulting prior-data conflict is less strong under the skeptical prior than
under the skeptical mixture.
As a consequence, we now obtain
$BF_S=0.15<BF_{SM}=0.36$ so that evidence for replication success is moderate under the skeptical and only anecdotal under the skeptical mixture; a similar phenomenon holds for the study \cite{morewedge2010thought}. Clearly for both studies by setting a higher value of $\alpha$ such as 0.05 or  0.1 we obtain agreement.

%

\black
\renewcommand{\arraystretch}{0.5}
\begin{table}
\begin{small}
\caption{\tiny Twelve studies from the \emph{Social Sciences Replication Project} (Camerer et al., 2018).  Effect values, originally expressed as sample  correlation coefficients,   were subsequently turned into effect estimates  $\hat{\theta}$ using Fisher $z$-transformation.
Reported are the $z$-values   for the original ($z_o$) and replication studies ($z_r$); $c \approx n_r/n_o$, and
relative effect estimates $d = \hat{\theta}_r/\hat{\theta}_o$.
 Based on the choice $\gamma_S = BF_S$ and $\gamma_{SM} = BF_{SM}$ for skepticism, respectively, \color{black}
prior hyperparameters are  shown, namely
the relative variance $g_S$ for the skeptical prior and the  pair $(\psi_{SM, \alpha}, h_{SM, \alpha})$ for the skeptical mixture prior,
 the latter based on three possible prior-data conflict scenarios for $\alpha = \{ 0.01, 0.05, 0.1 \}$,  when these thresholds are achievable.
 \color{black}
$P_S(\hthetao)$ indicates the $p$-value for prior-data conflict under the skeptical prior, whereas $P_{SM}(\hthetao)$ indicates the $p$-value for prior-data conflict under the skeptical mixture prior.
  The skeptical Bayes factors $BF_{S}$, the replication Bayes factor $BF_R$, and the skeptical mixture Bayes factor  $BF_{SM}(\alpha)$ are  reported in the last three columns.
}
\label{tab:tabone}
\begin{center}
\begin{tabular}{ l@{\hspace{0.7\tabcolsep}}c@{\hspace{0.8\tabcolsep}}c@{\hspace{0.7\tabcolsep}}c@{\hspace{0.5\tabcolsep}}c@{\hspace{0.5\tabcolsep}}c@{\hspace{0.6\tabcolsep}}c@{\hspace{0.6\tabcolsep}}c@{\hspace{0.6\tabcolsep}} c@{\hspace{0.5\tabcolsep}} c@{\hspace{0.5\tabcolsep}}c@{\hspace{0.3\tabcolsep}} c@{\hspace{0.3\tabcolsep}} c@{\hspace{0.3\tabcolsep}} c@{\hspace{0.3\tabcolsep}} c@{\hspace{0.3\tabcolsep}}}
\hline
 \textbf{Study} & $z_o$ & $z_r$ & $n_o$ & $n_r$&  $c$ & $d$ & $g_S$& $P_S$ &$P_{SM}$ & $\psi_{SM, \alpha}$& $h_{SM, \alpha}$ & $BF_{S}$ & $BF_R$  & $BF_{SM}$ \\
\hline
\bm{$\alpha =0.01$} & & & & & & & & & & & & & &  \\
 & & & & & & & & & & & & & &  \\
 Aviezer et al.  & 6.8 & 3.93 & 15 & 14 & 0.92 & 0.6 & 0.24 & \tiny $<0.001$ & \tiny $<0.001$ &\tiny $<0.001$ & 0.24 & 0.013 & \tiny $<0.001$ & 0.013 \\
   Balaf. and S. & 2.37 & 2.28 & 72 & 243 & 3.48 & 0.52 & 0.25 & 0.034 & 0.034 & \tiny $<0.001$ & 0.25 & 0.638 & 0.26 & 0.638 \\
   Derex et al. & 4.04 & 2.97 & 51 & 65 & 1.29 & 0.65 & 0.4 & 0.001 & 0.01 & 0.963 & 12.26 & 0.117 & 0.03 & 0.05 \\
   Duncan et al. & 2.83 & 4.41 & 15 & 92 & 7.42 & 0.57 & 0.5 & 0.021 & 0.01 & 0.303 & 0.21 & 0.322 & \tiny $<0.001$  & 0.545 \\
   Gneezy et al. & 3 & 3.71 & 178 & 407 & 2.31 & 0.81 & 1 & 0.034 & 0.01 & 0.173 & 0.38 & 0.149 & \tiny $<0.001$ & 0.356 \\
   Janssen et al. & 5.76 & 2.24 & 63 & 42 & 0.65 & 0.48 & 0.03 & \tiny $<0.001$ & \tiny $<0.001$ & \tiny $<0.001$ & 0.03 & 0.63 & 0.61 & 0.63 \\
    Karp. and Bl. & 4.24 & 2.75 & 40 & 49 & 1.24 & 0.58 & 0.26 & \tiny $<0.001$ & 0.01 & 0.972 & 20 & 0.179 & 0.08 & 0.031 \\
   Kovacs et al. & 2.22 & 6.44 & 24 & 95 & 4.38 & 1.38 & 3.95 & 0.317 & 0.317 & \tiny $<0.001$ & 3.95 & 0.309 & \tiny $<0.001$ & 0.309 \\
   Morewedge et al. & 2.63 & 3.44 & 32 & 89 & 2.97 & 0.76 & 0.97 & 0.061 & 0.01 & 0.809 & 0.05 & 0.256 & 0.01 & 0.874 \\
   Nishi et al. & 2.85 & 2.55 & 200 & 480 & 2.42 & 0.57 & 0.35 & 0.014 & 0.011 & 0.189 & 0.32 & 0.401 & 0.12 & 0.529 \\
   Pyc and Rawson & 2.27 & 2.63 & 36 & 306 & 9.18 & 0.38 & 0.09 & 0.029 & 0.029 & \tiny $<0.001$ & 0.09 & 0.849 & 0.25 & 0.849 \\
   Rand et al. & 2.62 & 1.19 & 343 & 2136 & 6.27 & 0.18 & - &-  & - & \tiny $<0.001$ &-  & - & 9.59 & - \\
   & & & & & & & & & & & & & &  \\

   \bm{$\alpha =0.05$} & & & & & & & & & & & & & &  \\
 & & & & & & & & & & & & & &  \\

    Aviezer et al.  & 6.8 & 3.93 & 15 & 14 & 0.92 & 0.6 & 0.24 & \tiny $<0.001$ & \tiny $<0.001$ & \tiny $<0.001$ & 0.24 & 0.013 & \tiny $<0.001$ & 0.013 \\
   Balaf. and S. & 2.37 & 2.28 & 72 & 243 & 3.48 & 0.52 & 0.25 & 0.034 & 0.05 & 0.561 & 0.97 & 0.638 & 0.26 & 0.553 \\
   Derex et al. & 4.04 & 2.97 & 51 & 65 & 1.29 & 0.65 & 0.4 & 0.001 & 0.05 & 0.927 & 100 & 0.117 & 0.03 & 0.042 \\
   Duncan et al. & 2.83 & 4.41 & 15 & 92 & 7.42 & 0.57 & 0.5 & 0.021 & 0.05 & 0.663 & 2.67 & 0.322 & \tiny $<0.001$ & 0.256 \\
   Gneezy et al. & 3 & 3.71 & 178 & 407 & 2.31 & 0.81 & 1 & 0.034 & 0.05 & 0.458 & 2.13 & 0.149 & \tiny $<0.001$ & 0.143 \\
   Janssen et al. & 5.76 & 2.24 & 63 & 42 & 0.65 & 0.48 & 0.03 & \tiny $<0.001$ & \tiny $<0.001$ & \tiny $<0.001$ & 0.03 & 0.63 & 0.61 & 0.63 \\
   Karp. and Bl. & 4.24 & 2.75 & 40 & 49 & 1.24 & 0.58 & 0.26 & \tiny $<0.001$ & \tiny $<0.001$ & \tiny $<0.001$ & 0.26 & 0.179 & 0.08 & 0.179 \\
   Kovacs et al. & 2.22 & 6.44 & 24 & 95 & 4.38 & 1.38 & 3.95 & 0.317 & 0.05 & 0.009 & 0.31 & 0.309 & \tiny $<0.001$ & 0.653 \\
   Morewedge et al. & 2.63 & 3.44 & 32 & 89 & 2.97 & 0.76 & 0.97 & 0.061 & 0.05 & 0.012 & 0.81 & 0.256 & 0.01 & 0.287 \\
   Nishi et al. & 2.85 & 2.55 & 200 & 480 & 2.42 & 0.57 & 0.35 & 0.014 & 0.05 & 0.744 & 3.59 & 0.401 & 0.12 & 0.275 \\
   Pyc and Rawson & 2.27 & 2.63 & 36 & 306 & 9.18 & 0.38 & 0.09 & 0.029 & 0.05 & 0.71 & 1.1 & 0.849 & 0.25 & 0.674 \\
   Rand et al. & 2.62 & 1.19 & 343 & 2136 & 6.27 & 0.18 & - & -  & - & 0.948 &-  & - & 9.59 & - \\
   & & & & & & & & & & & & & &  \\

\bm{$\alpha =0.1$} & & & & & & & & & & & & & &  \\
 & & & & & & & & & & & & & &  \\
   Aviezer et al.  & 6.8 & 3.93 & 15 & 14 & 0.92 & 0.6 & 0.24 & \tiny $<0.001$ & \tiny $<0.001$ & \tiny $<0.001$ & 0.24 & 0.013 & \tiny $<0.001$ & 0.013 \\
   Balaf. and S. & 2.37 & 2.28 & 72 & 243 & 3.48 & 0.52 & 0.25 & 0.034 & 0.1 & 0.634 & 3.11 & 0.638 & 0.26 & 0.466 \\
   Derex et al. & 4.04 & 2.97 & 51 & 65 & 1.29 & 0.65 & 0.4 & 0.001 & 0.001 & \tiny $<0.001$ & 0.4 & 0.117 & 0.03 & 0.117 \\
   Duncan et al. & 2.83 & 4.41 & 15 & 92 & 7.42 & 0.57 & 0.5 & 0.021 & 0.1 & 0.667 & 6.2 & 0.322 & \tiny $<0.001$ & 0.219 \\
   Gneezy et al. & 3 & 3.71 & 178 & 407 & 2.31 & 0.81 & 1 & 0.034 & 0.1 & 0.608 & 5.82 & 0.149 & \tiny $<0.001$ & 0.132 \\
   Janssen et al. & 5.76 & 2.24 & 63 & 42 & 0.65 & 0.48 & 0.03 & \tiny $<0.001$ & \tiny $<0.001$ & \tiny $<0.001$ & 0.03 & 0.63 & 0.61 & 0.63 \\
   Karp. and Bl. & 4.24 & 2.75 & 40 & 49 & 1.24 & 0.58 & 0.26 & \tiny $<0.001$ & \tiny $<0.001$ & \tiny $<0.001$ & 0.26 & 0.179 & 0.08 & 0.179 \\
   Kovacs et al. & 2.22 & 6.44 & 24 & 95 & 4.38 & 1.38 & 3.95 & 0.317 & 0.1 & 0.002 & 1.7 & 0.309 & \tiny $<0.001$ & 0.441 \\
   Morewedge et al. & 2.63 & 3.44 & 32 & 89 & 2.97 & 0.76 & 0.97 & 0.061 & 0.1 & 0.406 & 2.55 & 0.256 & 0.01 & 0.238 \\
   Nishi et al. & 2.85 & 2.55 & 200 & 480 & 2.42 & 0.57 & 0.35 & 0.014 & 0.1 & 0.73 & 8.65 & 0.401 & 0.12 & 0.245 \\
   Pyc and Rawson & 2.27 & 2.63 & 36 & 306 & 9.18 & 0.38 & 0.09 & 0.029 & 0.1 & 0.688 & 3.26 & 0.849 & 0.25 & 0.561 \\
   Rand et al. & 2.62 & 1.19 & 343 & 2136 & 6.27 & 0.18 & - &-  & - & 0.884 &-  & -  & 9.59 &-  \\

  	\hline
\end{tabular}
\end{center}
\end{small}
\end{table}


\begin{center}
INCLUDE HERE Figure \ref{fig:fourth}
\end{center}

\begin{center}
INCLUDE HERE Figure \ref{fig:fifth}
\end{center}

\begin{center}
INCLUDE HERE Figure \ref{fig:sixth}
\end{center}

\section{Discussion}
\label{sec:discussion}



\color{black}

We  presented a method to analyze replication studies and in particular to assess whether a replication experiment  is successful in reproducing the findings of  an original experiment.
We used a stylized framework with effect size estimators approximately normally distributed with known variances, a procedure often used in meta-analysis, possibly after a suitable transformation,  with reasonably large sample sizes. 

Throughout we systematically used  the  Bayes factor (BF) as a measure of evidence, coupled with reverse-Bayes techniques to elicit a skeptical prior, along the lines originally presented in \citet{pawel2022sceptical}.
   We proposed a novel skeptical mixture prior which combines skepticism while hedging against prior-data conflict, a feature which is not currently available in the  methodology for skeptical BF's. For this reason, our  method   could be more attractive to external agencies or reviewers when evaluating a replication protocol,  in particular when a standard skeptical prior is strongly in conflict with the original data.
In this context, a useful  take-home message of our investigation is that
restoring prior-data conflict to
a tolerable level  may  lead to declare stronger evidence for replication success,
as illustrated in the analysis of  some cases within the Social Science Replication Project.

We evaluated prior-data conflict using the notion of $p$-value proposed by \citet{evans2006checking}
and examined  sensitivity of our results to a grid of $\alpha$-values.
However our framework can be employed with alternative measures of conflict, such as those presented in
\citet{reimherr-emg-nicolae-2021} or  in   \cite{young1996measuring} and \cite{veen2018using}.

\bigskip
\begin{center}
{\large\bf ACKNOWLEDGMENTS}
\end{center}

We express our thanks to Leonhard Held   and Samuel Pawel (University of Zurich) for useful discussions on the issue of  replication studies.
\black

\bigskip
\begin{center}
{\large\bf SUPPLEMENTARY MATERIAL}
\end{center}

The Supplementary material contains  technical results.
Specifically: the proofs of Proposition 1 and 2 and the proof of information consistency for the replication BF.

Part of the data appearing in Table \ref{tab:tabone} was downloaded from  Samuel Pawel's Github page: \url{https://github.com/SamCH93/ReplicationSuccess}.

\bigskip
\begin{center}
{\large\bf FUNDING DETAILS}
\end{center}

Partial financial support for GC was provided by UCSC, projects D1 years 2021-2023.

\black

\bigskip
\begin{center}
{\large\bf 	DISCLOSURE STATEMENT}
\end{center}
The authors report that there are no competing interests to declare. A Conflict of Interest statement is included at the end of the manuscript.

\bigskip
\begin{center}
{\large\bf 	CONFLICT OF INTEREST}
\end{center}
None.

\bibliographystyle{Chicago}

\begin{thebibliography}{}

\bibitem[\protect\citeauthoryear{Anderson and Maxwell}{Anderson and
  Maxwell}{2016}]{anderson2016there}
Anderson, S.~F. and S.~E. Maxwell (2016).
\newblock There’s more than one way to conduct a replication study: Beyond
  statistical significance.
\newblock {\em Psychological Methods\/}~{\em 21\/}(1), 1--12.

\bibitem[\protect\citeauthoryear{Aviezer, Trope, and Todorov}{Aviezer
  et~al.}{2012}]{aviezer2012body}
Aviezer, H., Y.~Trope, and A.~Todorov (2012).
\newblock Body cues, not facial expressions, discriminate between intense
  positive and negative emotions.
\newblock {\em Science\/}~{\em 338\/}(6111), 1225--1229.

\bibitem[\protect\citeauthoryear{Balafoutas and Sutter}{Balafoutas and
  Sutter}{2012}]{balafoutas2012affirmative}
Balafoutas, L. and M.~Sutter (2012).
\newblock Affirmative action policies promote women and do not harm efficiency
  in the laboratory.
\newblock {\em Science\/}~{\em 335\/}(6068), 579--582.

\bibitem[\protect\citeauthoryear{Bayarri, Berger, Forte, and
  Garc{\'\i}a-Donato}{Bayarri et~al.}{2012}]{bayarri2012criteria}
Bayarri, M.~J., J.~O. Berger, A.~Forte, and G.~Garc{\'\i}a-Donato (2012).
\newblock Criteria for {B}ayesian model choice with application to variable
  selection.
\newblock {\em The Annals of {S}tatistics\/}~{\em 40\/}(3), 1550--1577.

\bibitem[\protect\citeauthoryear{Bernardo and Smith}{Bernardo and
  Smith}{2000}]{Bernardo:Smith:2000}
Bernardo, J.~M. and A.~Smith (2000).
\newblock {\em Bayesian Theory}.
\newblock Wiley series in probability and statistics. Chichester: John Wiley
  and Sons Ltd.

\bibitem[\protect\citeauthoryear{Camerer, Dreber, Holzmeister, Ho, Huber,
  Johannesson, Kirchler, Nave, Nosek, Pfeiffer, et~al.}{Camerer
  et~al.}{2018}]{camerer2018evaluating}
Camerer, C.~F., A.~Dreber, F.~Holzmeister, T.-H. Ho, J.~Huber, M.~Johannesson,
  M.~Kirchler, G.~Nave, B.~A. Nosek, T.~Pfeiffer, et~al. (2018).
\newblock Evaluating the replicability of social science experiments in nature
  and science between 2010 and 2015.
\newblock {\em Nature Human Behaviour\/}~{\em 2\/}(9), 637--644.

\bibitem[\protect\citeauthoryear{Dawid}{Dawid}{2011}]{Dawid:2011}
Dawid, A.~P. (2011).
\newblock Posterior model probabilities.
\newblock In P.~S. Bandyopadhyay and M.~Forster (Eds.), {\em Philosophy of
  Statistics}, pp.\  607--630. Elsevier, Amsterdam.

\bibitem[\protect\citeauthoryear{Derex, Beugin, Godelle, and Raymond}{Derex
  et~al.}{2013}]{derex2013experimental}
Derex, M., M.-P. Beugin, B.~Godelle, and M.~Raymond (2013).
\newblock Experimental evidence for the influence of group size on cultural
  complexity.
\newblock {\em Nature\/}~{\em 503\/}(7476), 389--391.

\bibitem[\protect\citeauthoryear{Egidi, Pauli, and Torelli}{Egidi
  et~al.}{2021}]{egidi2021avoiding}
Egidi, L., F.~Pauli, and N.~Torelli (2021).
\newblock Avoiding prior--data conflict in regression models via mixture
  priors.
\newblock {\em Canadian Journal of Statistics\/}~{\em 50\/}(2), 491--510.

\bibitem[\protect\citeauthoryear{Evans and Moshonov}{Evans and
  Moshonov}{2006}]{evans2006checking}
Evans, M. and H.~Moshonov (2006).
\newblock Checking for prior-data conflict.
\newblock {\em Bayesian {A}nalysis\/}~{\em 1\/}(4), 893--914.

\bibitem[\protect\citeauthoryear{Gneezy, Keenan, and Gneezy}{Gneezy
  et~al.}{2014}]{gneezy2014avoiding}
Gneezy, U., E.~A. Keenan, and A.~Gneezy (2014).
\newblock Avoiding overhead aversion in charity.
\newblock {\em Science\/}~{\em 346\/}(6209), 632--635.

\bibitem[\protect\citeauthoryear{Good}{Good}{1950}]{Good1950book}
Good, I.~J. (1950).
\newblock Probability and the weighing of evidence.
\newblock {\em Philosophy\/}~{\em 26\/}(97), 163--164.

\bibitem[\protect\citeauthoryear{Harms}{Harms}{2019}]{harms2019bayes}
Harms, C. (2019).
\newblock A {B}ayes factor for replications of {A}{N}{O}{V}{A} results.
\newblock {\em The American Statistician\/}~{\em 73\/}(4), 327--339.

\bibitem[\protect\citeauthoryear{Hedges and Schauer}{Hedges and
  Schauer}{2019}]{hedges2019statistical}
Hedges, L.~V. and J.~M. Schauer (2019).
\newblock Statistical analyses for studying replication: Meta-analytic
  perspectives.
\newblock {\em Psychological Methods\/}~{\em 24\/}(5), 557--570.

\bibitem[\protect\citeauthoryear{Held}{Held}{2020}]{held2020new}
Held, L. (2020).
\newblock A new standard for the analysis and design of replication studies.
\newblock {\em Journal of the Royal Statistical Society: Series A (Statistics
  in Society)\/}~{\em 183\/}(2), 431--448.

\bibitem[\protect\citeauthoryear{Held, Matthews, Ott, and Pawel}{Held
  et~al.}{2022}]{held-etal-2022reverse}
Held, L., R.~Matthews, M.~Ott, and S.~Pawel (2022).
\newblock Reverse-{B}ayes methods for evidence assessment and research
  synthesis.
\newblock {\em Research Synthesis Methods\/}~{\em 13\/}(3), 295--314.

\bibitem[\protect\citeauthoryear{Hutton, Diggle, Bird, Hennig, Longford,
  Mathur, Vander~Weele, Ioannidis, Chai, Dowe, et~al.}{Hutton
  et~al.}{2020}]{hutton2020discussion}
Hutton, J.~L., P.~J. Diggle, S.~M. Bird, C.~Hennig, N.~Longford, M.~B. Mathur,
  T.~J. Vander~Weele, J.~P. Ioannidis, C.~P. Chai, D.~L. Dowe, et~al. (2020).
\newblock Discussion on the meeting on ‘signs and sizes: understanding and
  replicating statistical findings’.
\newblock {\em Journal of the Royal Statistical Society. Series A (Statistics
  in Society)\/}~{\em 183\/}(2), 449--469.

\bibitem[\protect\citeauthoryear{Janssen, Holahan, Lee, and Ostrom}{Janssen
  et~al.}{2010}]{janssen2010lab}
Janssen, M.~A., R.~Holahan, A.~Lee, and E.~Ostrom (2010).
\newblock Lab experiments for the study of social-ecological systems.
\newblock {\em Science\/}~{\em 328\/}(5978), 613--617.

\bibitem[\protect\citeauthoryear{Jeffreys}{Jeffreys}{1961}]{Jeffreys:1961}
Jeffreys, H. (1961).
\newblock {\em Theory of Probability (3rd Edition)}.
\newblock Oxford, University Press.

\bibitem[\protect\citeauthoryear{Johnson, Payne, Wang, Asher, and
  Mandal}{Johnson et~al.}{2017}]{johnson:etal:2017:On:the:Reproducibility}
Johnson, V.~E., R.~D. Payne, T.~Wang, A.~Asher, and S.~Mandal (2017).
\newblock On the reproducibility of psychological science.
\newblock {\em Journal of the American Statistical Association\/}~{\em
  112\/}(517), 1--10.

\bibitem[\protect\citeauthoryear{Karpicke and Blunt}{Karpicke and
  Blunt}{2011}]{karpicke2011retrieval}
Karpicke, J.~D. and J.~R. Blunt (2011).
\newblock Retrieval practice produces more learning than elaborative studying
  with concept mapping.
\newblock {\em Science\/}~{\em 331\/}(6018), 772--775.

\bibitem[\protect\citeauthoryear{Kass and Raftery}{Kass and
  Raftery}{1995}]{Kass:Raftery:1995}
Kass, R.~E. and A.~E. Raftery (1995).
\newblock Bayes factors.
\newblock {\em Journal of the American Statistical Association\/}~{\em
  90\/}(430), 773--795.

\bibitem[\protect\citeauthoryear{Kov{\'a}cs, T{\'e}gl{\'a}s, and
  Endress}{Kov{\'a}cs et~al.}{2010}]{kovacs2010social}
Kov{\'a}cs, {\'A}.~M., E.~T{\'e}gl{\'a}s, and A.~D. Endress (2010).
\newblock The social sense: Susceptibility to others’ beliefs in human
  infants and adults.
\newblock {\em Science\/}~{\em 330\/}(6012), 1830--1834.

\bibitem[\protect\citeauthoryear{Liang, Paulo, Molina, Clyde, and Berger}{Liang
  et~al.}{2008}]{liang2008mixtures}
Liang, F., R.~Paulo, G.~Molina, M.~A. Clyde, and J.~O. Berger (2008).
\newblock Mixtures of $g$-priors for {B}ayesian variable selection.
\newblock {\em Journal of the American Statistical Association\/}~{\em
  103\/}(481), 410--423.

\bibitem[\protect\citeauthoryear{Ly, Etz, Marsman, and Wagenmakers}{Ly
  et~al.}{2018}]{Ly:etal:2018:replication:bayes:factor}
Ly, A., A.~Etz, M.~Marsman, and E.-J. Wagenmakers (2018).
\newblock Replication {B}ayes factors from evidence updating.
\newblock {\em Behavior Research Methods\/}~{\em 51}, 2498--2508.

\bibitem[\protect\citeauthoryear{Ly, Etz, Marsman, and Wagenmakers}{Ly
  et~al.}{2019}]{ly2019replication}
Ly, A., A.~Etz, M.~Marsman, and E.-J. Wagenmakers (2019).
\newblock Replication bayes factors from evidence updating.
\newblock {\em Behavior research methods\/}~{\em 51}, 2498--2508.

\bibitem[\protect\citeauthoryear{Ly and Wagenmakers}{Ly and
  Wagenmakers}{2022}]{ly-wagenmakers2022bayes}
Ly, A. and E.-J. Wagenmakers (2022).
\newblock Bayes factors for peri-null hypotheses.
\newblock {\em TEST\/}~{\em 31\/}(4), 1121--1142.

\bibitem[\protect\citeauthoryear{Morewedge, Huh, and Vosgerau}{Morewedge
  et~al.}{2010}]{morewedge2010thought}
Morewedge, C.~K., Y.~E. Huh, and J.~Vosgerau (2010).
\newblock Thought for food: Imagined consumption reduces actual consumption.
\newblock {\em Science\/}~{\em 330\/}(6010), 1530--1533.

\bibitem[\protect\citeauthoryear{Nishi, Shirado, Rand, and Christakis}{Nishi
  et~al.}{2015}]{nishi2015inequality}
Nishi, A., H.~Shirado, D.~G. Rand, and N.~A. Christakis (2015).
\newblock Inequality and visibility of wealth in experimental social networks.
\newblock {\em Nature\/}~{\em 526\/}(7573), 426--429.

\bibitem[\protect\citeauthoryear{O'Hagan and Forster}{O'Hagan and
  Forster}{2004}]{o2004kendall}
O'Hagan, A. and J.~J. Forster (2004).
\newblock {\em Kendall's advanced theory of statistics, volume 2B: Bayesian
  inference}, Volume~2.
\newblock Arnold.

\bibitem[\protect\citeauthoryear{{Open Science Collaboration}}{{Open Science
  Collaboration}}{2015}]{open2015estimating}
{Open Science Collaboration} (2015).
\newblock Estimating the reproducibility of psychological science.
\newblock {\em Science\/}~{\em 349\/}(6251), aac4716.

\bibitem[\protect\citeauthoryear{Pawel and Held}{Pawel and
  Held}{2022}]{pawel2022sceptical}
Pawel, S. and L.~Held (2022).
\newblock The sceptical {B}ayes factor for the assessment of replication
  success.
\newblock {\em Journal of the Royal Statistical Soiety: Series B\/}~{\em
  84\/}(3), 879--911.

\bibitem[\protect\citeauthoryear{Pyc and Rawson}{Pyc and
  Rawson}{2010}]{pyc2010testing}
Pyc, M.~A. and K.~A. Rawson (2010).
\newblock Why testing improves memory: Mediator effectiveness hypothesis.
\newblock {\em Science\/}~{\em 330\/}(6002), 335--335.

\bibitem[\protect\citeauthoryear{Reimherr, Meng, and Nicolae}{Reimherr
  et~al.}{2021}]{reimherr-emg-nicolae-2021}
Reimherr, M., X.-L. Meng, and D.~L. Nicolae (2021).
\newblock Prior sample size extensions for assessing prior impact and
  prior-likelihood discordance.
\newblock {\em Journal of the Royal Statistical Society: Series B (Statistical
  Methodology)\/}~{\em 83\/}(3), 413--437.

\bibitem[\protect\citeauthoryear{Ro\v{c}kov\'{a}}{Ro\v{c}kov\'{a}}{2018}]{rockova2018bayesianestimation}
Ro\v{c}kov\'{a}, V. (2018).
\newblock {Bayesian estimation of sparse signals with a continuous
  spike-and-slab prior}.
\newblock {\em The Annals of Statistics\/}~{\em 46\/}(1), 401 -- 437.

\bibitem[\protect\citeauthoryear{Sch{\"o}nbrodt and Wagenmakers}{Sch{\"o}nbrodt
  and Wagenmakers}{2018}]{schonbrodt2018bayes}
Sch{\"o}nbrodt, F.~D. and E.-J. Wagenmakers (2018).
\newblock Bayes factor design analysis: Planning for compelling evidence.
\newblock {\em Psychonomic {B}ulletin \& {R}eview\/}~{\em 25\/}(1), 128--142.

\bibitem[\protect\citeauthoryear{Shafer}{Shafer}{1982}]{Shafer1982Lindleysparadox}
Shafer, G. (1982).
\newblock Lindley's paradox.
\newblock {\em Journal of the American Statistical Association\/}~{\em
  77\/}(378), 325--334.

\bibitem[\protect\citeauthoryear{Spiegelhalter, Abrams, and
  Myles}{Spiegelhalter et~al.}{2003}]{Spiegelhalter:et:al:book:2003}
Spiegelhalter, D., K.~Abrams, and J.~Myles (2003).
\newblock {\em Bayesian Approaches to Clinical Trials and Health-Care
  Evaluation}.
\newblock John Wiley \& Sons, Ltd.

\bibitem[\protect\citeauthoryear{Taylor and Pollard}{Taylor and
  Pollard}{2009}]{TaylorPollard2009hypothesistests}
Taylor, S. and K.~Pollard (2009).
\newblock Hypothesis tests for point-mass mixture data with application to
  {O}mics data with many zero values.
\newblock {\em Statistical Applications in Genetics and Molecular
  Biology\/}~{\em 8\/}(1), 1--43.

\bibitem[\protect\citeauthoryear{Veen, Stoel, Schalken, Mulder, and Van~de
  Schoot}{Veen et~al.}{2018}]{veen2018using}
Veen, D., D.~Stoel, N.~Schalken, K.~Mulder, and R.~Van~de Schoot (2018).
\newblock Using the data agreement criterion to rank experts’ beliefs.
\newblock {\em Entropy\/}~{\em 20\/}(8), 592.

\bibitem[\protect\citeauthoryear{Verhagen and Wagenmakers}{Verhagen and
  Wagenmakers}{2014}]{verhagen2014bayesian}
Verhagen, J. and E.-J. Wagenmakers (2014).
\newblock Bayesian tests to quantify the result of a replication attempt.
\newblock {\em Journal of Experimental Psychology: General\/}~{\em 143\/}(4),
  1457--1475.

\bibitem[\protect\citeauthoryear{Wetzels and Wagenmakers}{Wetzels and
  Wagenmakers}{2012}]{wetzels2012default}
Wetzels, R. and E.-J. Wagenmakers (2012).
\newblock A default {B}ayesian hypothesis test for correlations and partial
  correlations.
\newblock {\em Psychonomic {B}ulletin \& {R}eview\/}~{\em 19\/}(6), 1057--1064.

\bibitem[\protect\citeauthoryear{Young and Pettit}{Young and
  Pettit}{1996}]{young1996measuring}
Young, K. and L.~Pettit (1996).
\newblock Measuring discordancy between prior and data.
\newblock {\em Journal of the Royal Statistical Society: Series B
  (Methodological)\/}~{\em 58\/}(4), 679--689.

\end{thebibliography}

\newpage


\end{document}